\documentclass[final,5p,times,twocolumn]{elsarticle}

\usepackage{siunitx}
\usepackage{float}
\usepackage{amsmath}
\usepackage{eurosym}
\usepackage{tikz}
\usepackage{tikzscale}
\usetikzlibrary{calc}
\usepackage[hyphens]{url}
\usepackage{amssymb}
\usepackage{relsize}    
\usepackage{xspace}     

\journal{Nuclear Instruments and Methods in Physics Research A}

\begin{document}

\begin{frontmatter}



\title{The HiSPARC Experiment}


\author[nikhef]{K. van Dam\corref{cor}}
\ead{kaspervd@nikhef.nl}
\author[nikhef,twente]{B. van Eijk\corref{cor}}
\ead{vaneijk@nikhef.nl}
\author[vu]{D.B.R.A. Fokkema}
\author[nikhef,leiden]{J.W. van Holten}
\author[nikhef,industry]{A.P.L.S. de Laat}
\author[nikhef,zaanlands]{N.G. Schultheiss}
\author[nikhef]{J.J.M. Steijger}
\author[nikhef]{J.C. Verkooijen}

\address[nikhef]{Nikhef National Institute for Subatomic Physics, Science Park 105, 1098 XG Amsterdam, The Netherlands}
\address[twente]{Faculty of Science and Technology, University of Twente, 7500 AE Enschede, The Netherlands}
\address[vu]{Faculty of Sciences, VU University Amsterdam, De Boelelaan 1081, 1081 HV Amsterdam, The Netherlands}
\address[leiden]{Lorentz Institute, Leiden University, Niels Bohrweg 2A, 2333 CA Leiden, The Netherlands}
\address[zaanlands]{Zaanlands Lyceum, Vincent van Goghweg 42, 1506 JD Zaandam, The Netherlands}
\address[industry]{Now in industry}

\cortext[cor]{Corresponding authors}

\begin{abstract}
The High School Project on Astrophysics Research with Cosmics (HiSPARC) is a large extensive air shower (EAS) array with detection stations throughout the Netherlands, United Kingdom, Denmark and Namibia. HiSPARC is a collaboration of universities, scientific institutes and high schools. The majority of detection stations is hosted by high schools. A HiSPARC station consists of two or four scintillators placed inside roof boxes on top of a building. The measured response of a detector to single incoming muons agrees well with GEANT4 simulations. The response of a station to EASs agrees with simulations as well. A four-scintillator station was integrated in the KASCADE experiment and was used to determine the accuracy of the shower direction reconstruction. Using simulations, the trigger efficiency of a station to detect a shower as function of both distance to the shower core and zenith angle was determined. The HiSPARC experiment is taking data since 2003. The number of stations ($\sim$140 in 2019) still increases. The project demonstrates that its approach is viable for educational purposes and that scientific data can be obtained in a collaboration with high school students and teachers.
\end{abstract}


\end{frontmatter}


\section{Introduction}
\label{sec:introduction}
Cosmic rays are energetic particles from space that hit the Earth's atmosphere at a rate of about 1000 per square meter per second \cite{gaisser2016}. They mainly consist of protons ($\sim$90\%) and $\alpha$-particles ($\sim$9\%). A very small fraction contains heavier, ionized nuclei \cite{gaisser2016}. Cosmic rays with energies up to about $10^{10}$ eV are predominantly produced by the Sun (solar wind \cite{hundhausen2012} and solar energetic particles \cite{sepreview2017}). Cosmic rays with energies between $10^{10}$ eV and $10^{18}$ eV are considered to be of galactic origin \cite{blasi2013}. These galactic cosmic rays are believed to predominantly originate from supernova remnants. Particles with energies beyond $10^{18}$ eV up to the extreme energy of $10^{20}$ eV stem from extra-galactic sources \cite{auger2017}. However, little is known about these sources and acceleration mechanisms. The flux of galactic cosmic rays decreases rapidly with energy ($\sim$$\textup{E}^{-2.7}$ above $10^{10}$ eV and drops to $\sim$$\textup{E}^{-3.1}$ beyond $3 \times 10^{15}$ eV). Thus, solar cosmic rays are many orders of magnitude more abundant than galactic cosmic rays. The flux of these galactic cosmic rays is in turn many times higher than that of extra-galactic cosmic rays. The cosmic ray rate above $10^{15}$ eV quickly drops to single events per square meter per year. At $10^{18} $ eV this rate drops to single events per square kilometer per year. This implies that space-based experiments focusing on these energy ranges suffer from low statistics.

When an energetic cosmic ray hits the Earth's atmosphere it will most likely collide with a nitrogen or oxygen nucleus. A large number of (energetic) secondary particles may be produced. These secondary particles will interact with other atmospheric nuclei. The multiplication process continues until the energy becomes insufficient for further particle production. The result of this mechanism is called an 'Extensive Air Shower' (EAS) \cite{grieder2010}.

The size of the footprint of an EAS at the surface of the Earth rises with the energy of the primary cosmic ray. Nevertheless, statistical fluctuations lead to differences in the total number of particles reaching ground level up to factors of ten. An EAS consists mainly of gamma rays, electrons (positrons) and, to a lesser extent, muons and hadrons. Below $\sim$$10^{13}$ eV, the shower leaves no footprint and only some remnants reach the ground. For cosmic ray energies below $\sim$$10^{11}$ eV only one or two muons reach the Earth's surface and no electrons, positrons or gamma rays are left. Because of the high flux of these low energy cosmic rays there is a large number of isolated minimum ionizing muons. We refer to these muons as the single muon background as they do not stem from EAS. The footprint of an EAS ranges from meters to several kilometers in diameter for perpendicular incident primaries. Since it is difficult to cover large footprints with a single detector, EASs are usually sampled by relatively small detectors arranged in arrays. By reconstructing the EAS, properties of the original cosmic ray can be derived. Examples of sampling experiments are KASCADE (13 m detector separation, 200 $\times$ 200 $\textup{m}^2$) \cite{kascade2003}, KASCADE-Grande (137 m detector separation, 700 $\times$ 700 $\textup{m}^2$) \cite{apel2010}, AGASA (1 km detector separation, 100 $\textup{km}^2$) \cite{agasa1992}, Telescope array (1.2 km detector separation, 762 $\textup{km}^2$) \cite{sdtelescopearray2012} and the Pierre Auger Observatory (1.5 km detector separation, 3000 $\textup{km}^2$) \cite{auger2015}.

\begin{figure}
\centering
\includegraphics[width=90 mm]{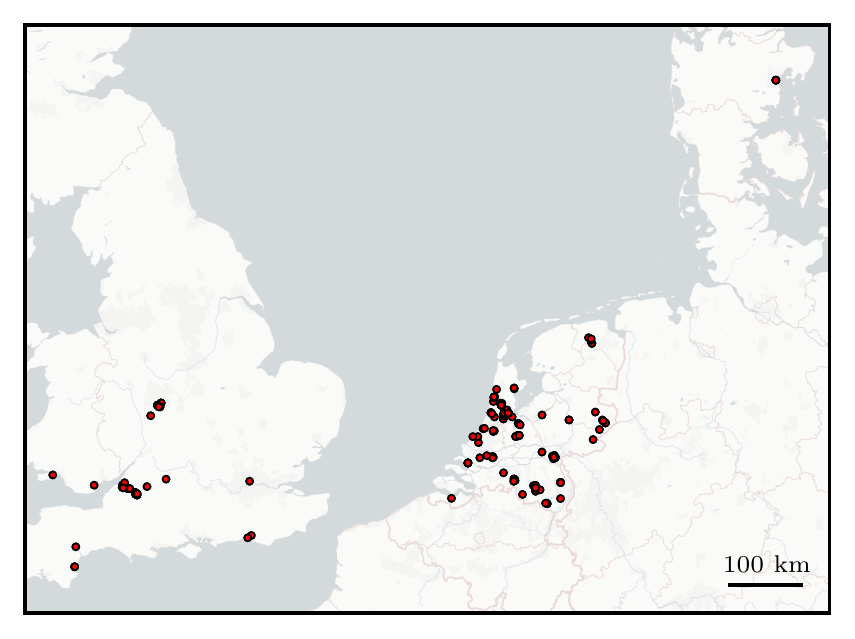}
\caption{Layout (early 2019) of the HiSPARC array. Each red dot represents one or more stations.}
\label{fig:hisparcarray}
\end{figure}

The High School Project on Astrophysics Research with Cosmics (HiSPARC) \cite{hisparcwebsite,proefschriftDavid} has approximately 140 detection stations distributed throughout the Netherlands, United Kingdom and Denmark (Fig. \ref{fig:hisparcarray}), and Namibia. HiSPARC is a collaboration of universities, scientific institutes and high schools each hosting their own detection station(s). The majority of the stations is located at high schools. HiSPARC has a strong outreach component. Stations are maintained by high school teachers, their students and university staff. Data are stored at Nikhef, the Dutch National Institute for Subatomic Physics \cite{nikhefwebsite}. A number of HiSPARC stations also employs a weather station. Their data are stored in the central database at Nikhef as well. Weather stations are used to analyse atmospheric conditions affecting EAS development.

Irregular arrays of cosmic ray detectors are not new, e.g. LAAS \cite{laas1999}, SEASA \cite{hofverberg2006}, CHICOS \cite{chicos2009}, and recently EEE \cite{eee2013}. The latter three projects have a strong education component involving high schools as well. In 2001, the Nijmegen Area High School Array (NAHSA \cite{timmermans2003}) was founded. In 2003 Nikhef initiated the expansion of NAHSA into a nationwide network: HiSPARC. The geometry of the array is determined by the location of the collaborating institutions. This has lead to an irregular grid of detection stations.

HiSPARC detection stations are relatively robust, cheap (5,000 \euro/10,000 \euro, configuration dependent), small and straightforward to assemble. Single (remote) stations are mainly used for educational purposes and to provide local measurements of the single muon flux, EAS flux and EAS directions. Clusters of stations can be used for more advanced EAS reconstruction. Typically, high schools have limited resources for both building, and maintaining a station. In this paper the HiSPARC experiment, its hardware infrastructure, data acquisition, analysis tools and performance are described.

\section{Scintillation detector}
\label{sec:detector}
The detection philosophy of HiSPARC is to sample EAS footprints using scintillation detectors. The light output of a scintillator is proportional to the number of charged EAS particles traversing the detector.

\begin{figure}
\centering
\includegraphics[width=90 mm]{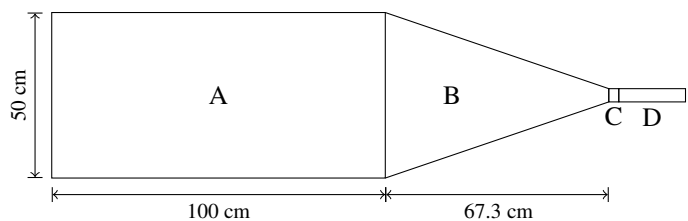}
\caption{Sketch of the HiSPARC detector. The scintillator and light-guide are denoted by the letters A and B resp. The light-guide adapter piece (C) enables the cylindrical PMT (D) to be mounted to the square end of the light-guide.}
\label{fig:detector}
\end{figure}

\begin{figure}
\centering
\includegraphics[width=90 mm]{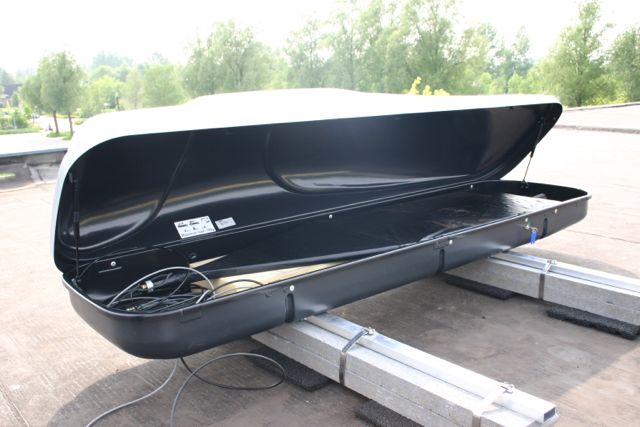}
\caption{The HiSPARC detector inside a roof box.}
\label{fig:photo}
\end{figure}

A scintillator (100 cm x 50 cm x 2 cm), see A in Figure \ref{fig:detector}, is glued to a slightly thicker triangular light-guide (base 50 cm, top 2.5 cm, height 67.5 cm, B) which is connected to a photomultiplier tube (PMT, D). A small adapter light-guide (C) connects to the cylindrical PMT. Both light-guide (polymethylmethacrylate or PMMA) and scintillator have comparable refractive indices (1.49 and 1.58 resp.). The scintillation material \cite{bc408datasheet} has a light attenuation length of 380 cm. The wavelength of maximum emission is 425 nm. The surfaces of the scintillator and light-guides are diamond polished to achieve a high surface reflectivity. The detector is wrapped in aluminum foil (thickness \SI{30}{\micro\meter}) and is made light-tight with black pond liner (thickness 0.45 mm). The assembly is placed inside a roof box (Fig. \ref{fig:photo}).

\subsection{Expected energy loss}
When a charged particle traverses the scintillator it loses energy. Apart from radiative energy loss, the particle primarily loses energy due to inelastic collisions with the atomic electrons of the scintillation material. Multiple molecules will be excited. These molecules will quickly return to their ground state and emit photons. The wavelength of the scintillation photons matches the wavelength characteristics of the PMT \cite{pmtelectrontubes, pmthamamatsu}. The photons may scatter several times inside the detector and only a fraction will reach the PMT. Figure \ref{fig:traceexmaple} shows a typical PMT output signal. The analog signal is sampled by an ADC in 2.5 ns bins. The area under the curve, or pulse integral, is a measure for the number of scintillation photons that have reached the PMT. The pulse integral is then calculated by summing the ADC values for each bin exceeding $-10$ mV (dotted line in Fig. \ref{fig:traceexmaple}).

The majority of muons and electrons (positrons) in an EAS carry high energy and can be considered minimum ionizing (MIP). For electron and muon MIPs the energy loss distribution is similar and is described by Landau theory \cite{landau1944}. The Landau distribution has a pronounced tail towards higher energy losses. A vertically incident minimum ionizing particle has a most probable energy loss in 2 cm scintillation material of 3.51 MeV ($\equiv$ 1 MIP).

Only a small fraction of the gamma rays in an EAS interacts with the scintillator via Compton scattering and less frequently, depending on their energy, via pair creation. These Compton electrons (positrons) are again detected via scintillation. Both direction and energy of the Compton electron depend on the scattering angle. The interaction depth differs for each gamma. Therefore, Compton electrons have different energies and travel different distances. The energy loss distribution due to gamma rays is a continuously decreasing function with energy. Pair creation is suppressed as high energy gamma rays are not abundant.

\subsection{PMT}
\begin{figure}
\centering
\includegraphics[width=90 mm]{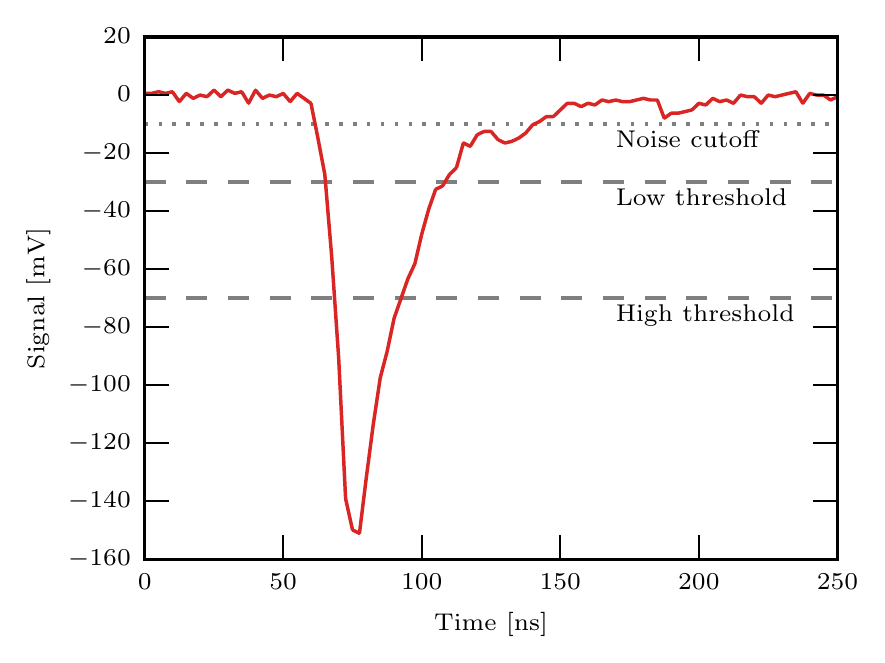}
\caption{Example of a typical signal with a pulse height of 150 mV. The FWHM is $\sim$25 ns. The pulse integral is calculated by summing all values in the bins where the signal exceeds $-10$ mV (dotted grey line). The default trigger thresholds are $-30$ and $-70$ mV (dashed grey lines).}
\label{fig:traceexmaple}
\end{figure}

\begin{figure}
\centering
\includegraphics[width=90 mm]{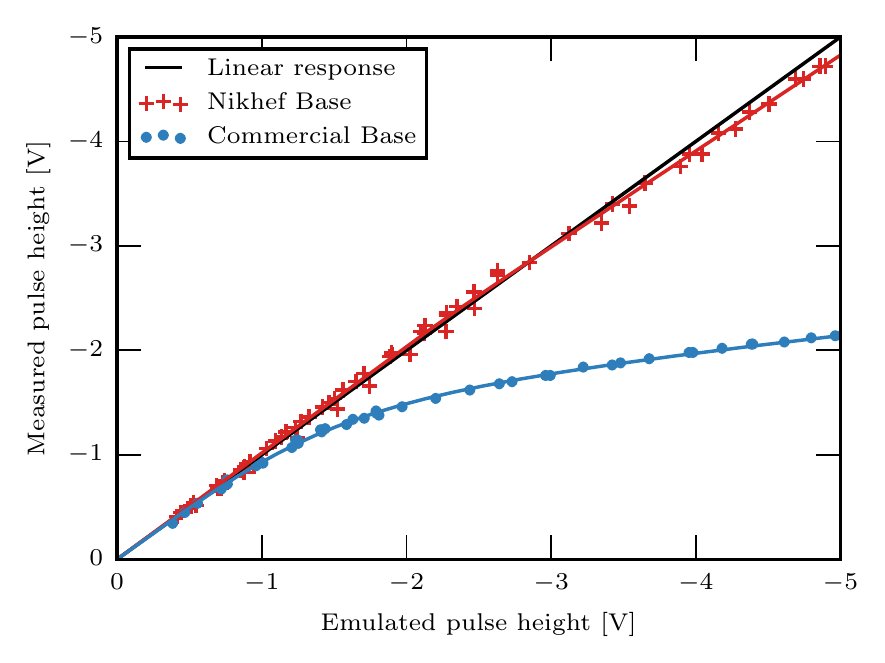}
\caption{Response of two HiSPARC PMT assemblies with different bases. On the horizontal axis the emulated pulse height (the combined pulse intensity of multiple LEDs) is shown. The vertical axis shows the measured pulse height (mV). The blue dots indicate the response for the commercial base whereas the red crosses show the pulse heights for the Nikhef base. Both response curves can be well described by the expression in eq. \ref{eq:pmtparam}. The black line indicates an ideal linear response. Up to about $-0.7$ V ($\sim$5 MIPs) both assemblies give a linear response.}
\label{fig:sensitivitypmt}
\end{figure}

HiSPARC deploys PMTs with a cathode diameter of 25 mm. The 12 cm long glass tube is enclosed by mu-metal, providing shielding against external magnetic fields. The PMT-base is supplied with a DC voltage of $-12$ V which is converted into a DC voltage ranging from $-300$ to $-1500$ V. The quantum efficiency at 425 nm is typically 25\%. Two different types of PMT bases are used. A commercial one \cite{baseelectrontubes} and an in-house developed version \cite{nikhefbase}. The Nikhef base provides a highly linear response over a large dynamical range, allowing to generate signals well in excess of $-5$ V (Fig. \ref{fig:sensitivitypmt}, red crosses). The rise-time is almost independent of the output signal amplitude. The impedance of the line driver, cable and subsequent readout electronics have been matched. The dynamical range of the commercial base is limited (Fig. \ref{fig:sensitivitypmt}, blue dots). The response curves were measured with a device containing 24 LEDs (light-emitting diodes). The LEDs were connected to the PMT using optical fibers. The pulsed light-output of each single LED was measured with the same PMT. Higher light intensities were obtained by bundling optical fibers. Figure \ref{fig:sensitivitypmt} shows that for small intensities both PMT assemblies behave linearly but at higher intensities the output flattens for the commercial base. The response of the PMT assemblies can be parametrised with a single function:
\begin{equation}
\label{eq:pmtparam}
f(x) = \frac{ax}{(x^{b}+c)^{\frac{1}{b}}}+dx
\end{equation}
with $a=0.237$, $b=13.5$, $c=9.34\cdot10^4$, $d=0.918$ for the Nikhef base and $a=1.42$, $b=2.74$, $c=4.13$, $d=0.150$ for the commercial base. A lab test on each PMT is not required. The PMT response function is derived from experimental data using the pulse integral (Fig. \ref{fig:traceexmaple}). Single minimum ionizing particles generate a peak in the pulse height and pulse integral distributions. The MIP-peak value in the pulse height domain ($\textup{MIP}_\textup{ph}$) is divided by the MIP-peak value in the pulse integral domain ($\textup{MIP}_\textup{pi}$). The pulse integrals ($\textup{pi}$) are subsequently multiplied by this ratio to obtain inferred pulse heights:
\begin{equation}
 \textup{ph}_\textup{inferred} = \frac{\textup{MIP}_\textup{ph}}{\textup{MIP}_\textup{pi}} \cdot \textup{pi}
\end{equation}
The measured pulse heights are expressed as function of the inferred pulse heights and the parametrisation in eq. (\ref{eq:pmtparam}) is fitted to obtain the PMT response function.

For perpendicular incident $10^{15}$ eV proton showers, electron densities (simulation) larger than \SI{50}{m^{-2}} occur within a \SI{\sim8}{\meter} radius. The high voltage on a PMT is chosen such that the MIP response distribution peaks at $\sim$$-150$ mV. The low and high thresholds are set to $-30$ mV (0.2 MIP) and $-70$ mV ($\sim$0.5 MIP) resp. These settings have been chosen to increase the sensitivity for gamma rays and low energy electrons while maintaining a decent dynamic range (ADC, $\sim$2.3 V or $\sim$15 MIPs). For larger pulses the read-out electronics are equipped with adjustable comparators in order to still be able to get an estimate of the pulse shape. For details see Section \ref{sec:electronics}.

As atmospheric conditions change, the temperature of the PMT assembly can vary between \SI{-30}{\celsius} on cold winter nights and +\SI{60}{\celsius} when the Sun heats the air in the roof box. Temperature differences affect the height of the signal pulse. For the commercial base a higher temperature results in a lower gain. The Nikhef base shows a higher gain at higher temperatures. To arrive at a proper measure for the pulse height, temperature variations need to be taken into account.

\begin{figure}
\centering
\includegraphics[width=90 mm]{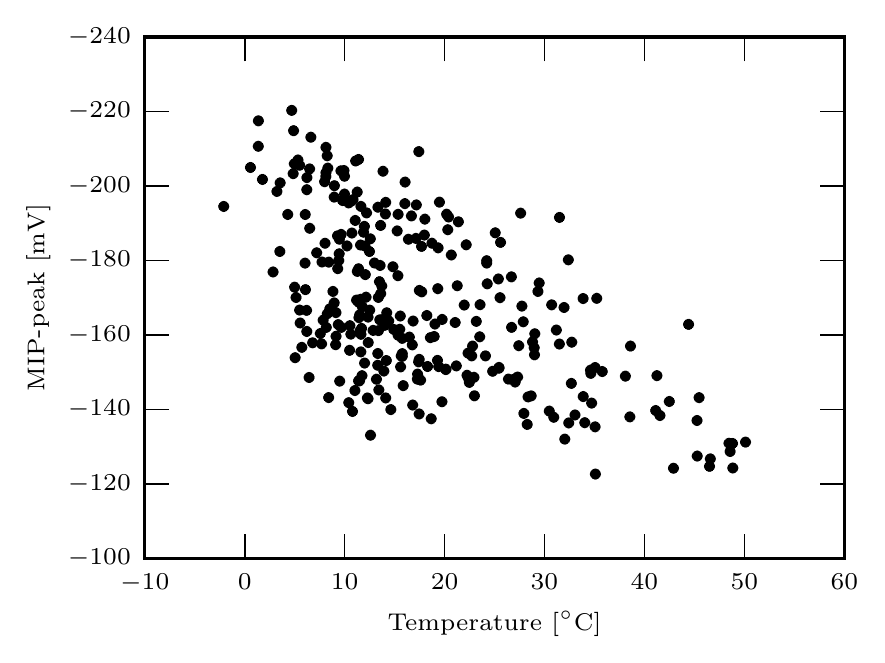}
\caption{Correlation between temperature and MIP-peak value for a PMT with a commercial base. The temperature was measured inside the roof box with a scale uncertainty of \SI{0.5}{\celsius}. The temperature and MIP-peak value were determined by averaging over 4-hour periods. There is a clear trend between the MIP-peak value and the local temperature.}
\label{fig:pmttemp}
\end{figure}

The value of the pulse height corresponding to 1 MIP is derived from EAS data as a function of temperature. The height of the MIP-peak is averaged over short periods in time (4 hours) and is used to calibrate the PMT output signal. Figure \ref{fig:pmttemp} gives an example of the correlation between temperature (measured inside the roof box with a scale uncertainty of \SI{0.5}{\celsius}) and MIP-peak value for a commercial assembly. Of order 1000 events are collected within 4 hours. A smaller time window results in a less accurate determination of the MIP-peak value. Alternatively, a method can be applied using a running average.

\subsection{Transmission of scintillation light}
When a (shower) particle propagates through the scintillator, the energy that is lost is converted into excitation photons. The yield of scintillation photons has been analysed using GEANT4 \cite{GEANT42016} by exposing the detector (scintillator and light-guides) to perpendicular incident minimum ionizing muons. As the scintillator has to be light tight, the detector is wrapped in black pond liner. To regain reflectivity, the scintillator is first packed in aluminum foil. The simulation assumes that the aluminum foil tightly encloses the detector. In reality, air pockets between scintillator and aluminum foil result in internal reflections. These reflections generate a higher yield of photons at the PMT than for a scintillator with a perfectly fitting aluminum envelope. In the simulation program this lack of air pockets is corrected for by increasing the aluminum reflectivity from 0.88 to a value of 0.93 to match the experimental data. This only scales the number of photons.

\begin{figure*}
\centering
\includegraphics[width=140mm]{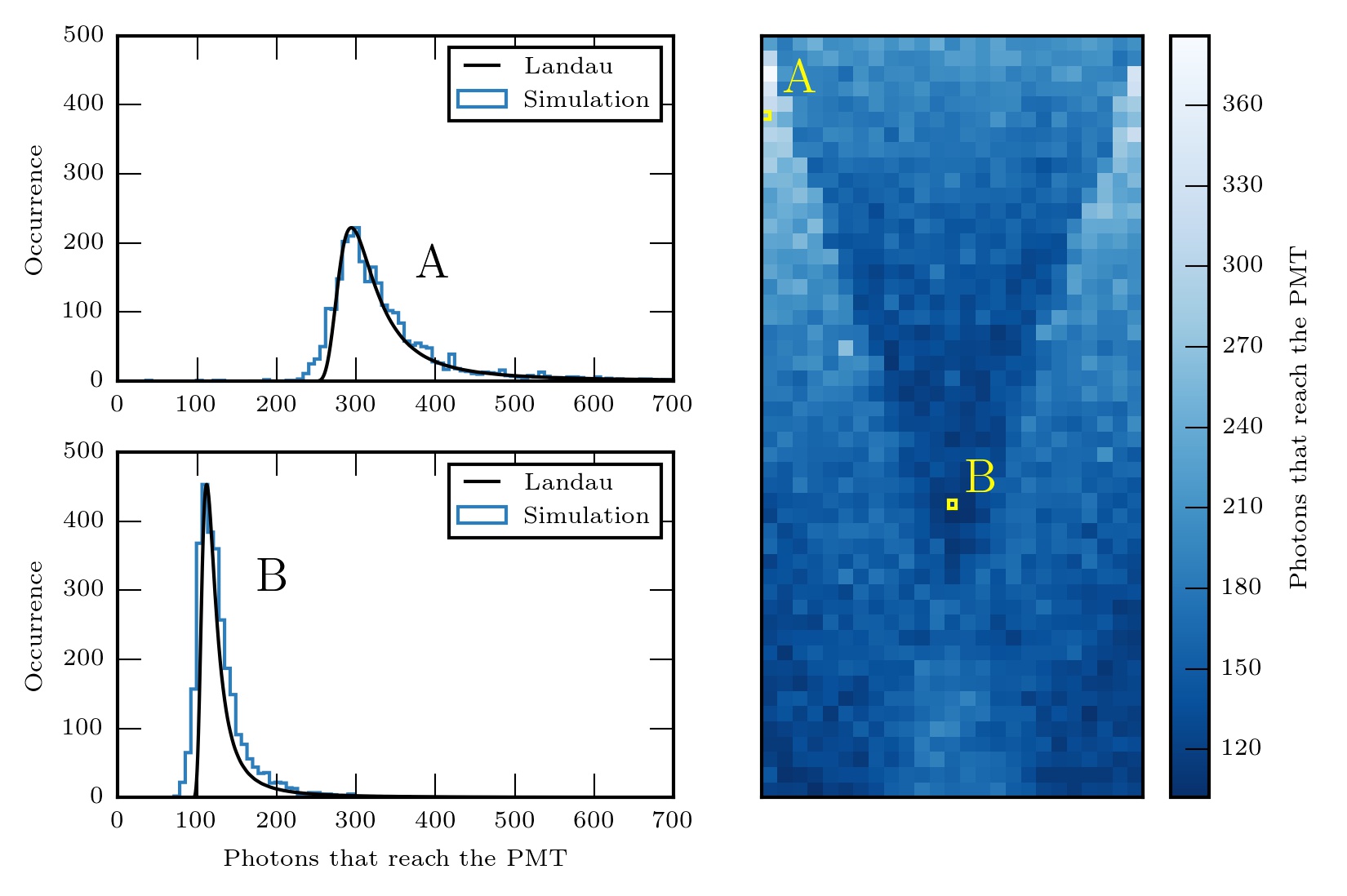}
\caption{At the right the average number of muon-induced scintillation photons arriving at the PMT as a function of the position in the detector is shown. In the simulation the detector was exposed to a large number of perpendicular incident relativistic muons. The muon flux was kept constant over the full area of the scintillator. The light-guide is connected at the top. The two histograms on the left show the fluctuation in the number of photons (in blue) released at locations A and B resp. Both distributions follow the Landau description (in black).}
\label{fig:sensitivity}
\end{figure*}

The photons have a different probability to reach the PMT depending on the location at which they are released in the scintillator. The plot on the right in Figure \ref{fig:sensitivity} shows the distribution of the number of scintillation photons arriving at the PMT as a function of position. The muon flux was kept constant over the full area of the detector. The left hand figures show the distribution of the number of photons that reach the PMT from locations A (top) and B (bottom) in the scintillator. The distribution of photons follows Landau theory (black curves). The discrepancy between simulation and Landau curves is believed to be due to local differences in attenuation and reflection.

Figure \ref{fig:sensitivity} also shows that the maximum light output is obtained when the muon hits a corner of the scintillator near the light-guide (top). Surprisingly, within the area that is represented by the mirror image of the light guide in the scintillator, the photon yield is relatively small.

\begin{figure}
\centering
\includegraphics[width=40mm]{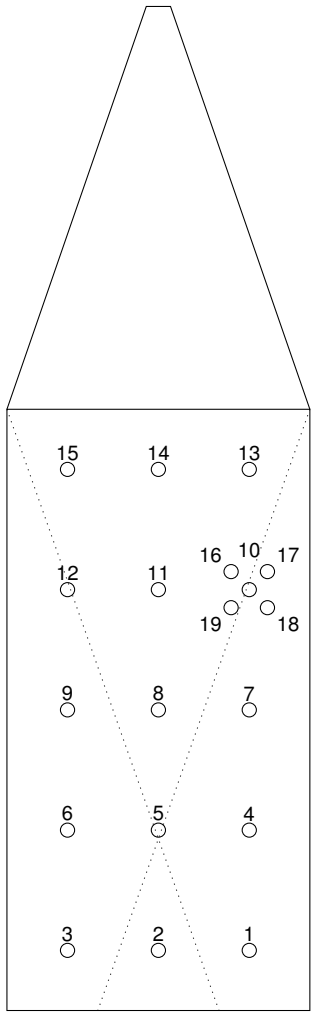}
\caption{The scintillator (including light-guide) light transmission is measured at 19 locations in the detector. The dotted lines show the position at which the largest gradients are observed (Fig. \ref{fig:sensitivity}). The circles indicate the positions at which the efficiency is measured. Fifteen positions are defined on a grid, with four additional measurements performed around point 10.}
\label{fig:transmission_experiment}
\end{figure}

\begin{figure}
\centering
\includegraphics[width=90mm]{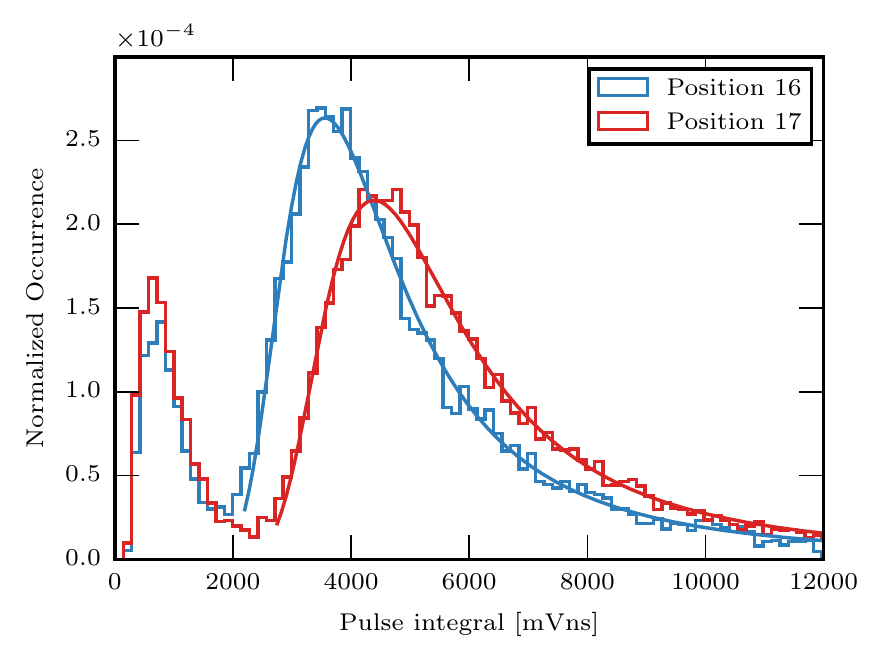}
\caption{Pulse integral distributions for events collected at position 16 and 17. Despite the small distance between the two positions there is a large difference in photon yield. Both distributions can accurately be described by a Landau convoluted with a Gaussian. The lower end of the spectrum results from gamma rays, low energy particles, and PMT noise.}
\label{fig:transmission_peaks}
\end{figure}

To verify the simulation a table-top experiment was designed \cite{proefschriftDavid}. A $\SI{1.5}{\centi\meter}\times\SI{1.5}{\centi\meter}$ scintillator is connected to a small PMT. This small scintillator is positioned on top of the HiSPARC detector. A $3\times 5$ grid (Fig. \ref{fig:transmission_experiment}, 1-15) defines the locations. In the vicinity of for instance point 10, the simulation predicts a large light yield gradient. In this region four additional measurements (16-19) were carried out. The readout is triggered when both scintillators signal a MIP. In Figure \ref{fig:transmission_peaks} the pulse integral distributions obtained at positions 16 and 17 are compared. Although the locations are very near, there is a sizeable difference in the photon yield reflected in the shift of the peak value. Position 17 has the peak at $\sim$4400 mVns whereas the yield at position 16 ($\sim$3600 mVns) is by $\sim$20\% smaller. Both distributions can accurately be described by a Landau distribution convoluted with a Gaussian. PMT response characteristics are not included in this simulation. Moreover, simulations only account for perpendicular incident muons whereas the experiment is susceptible to muons from all directions where both scintillators generate a sufficiently large signal. This results in Gaussian smearing.

\begin{figure}
\centering
\includegraphics[width=90mm]{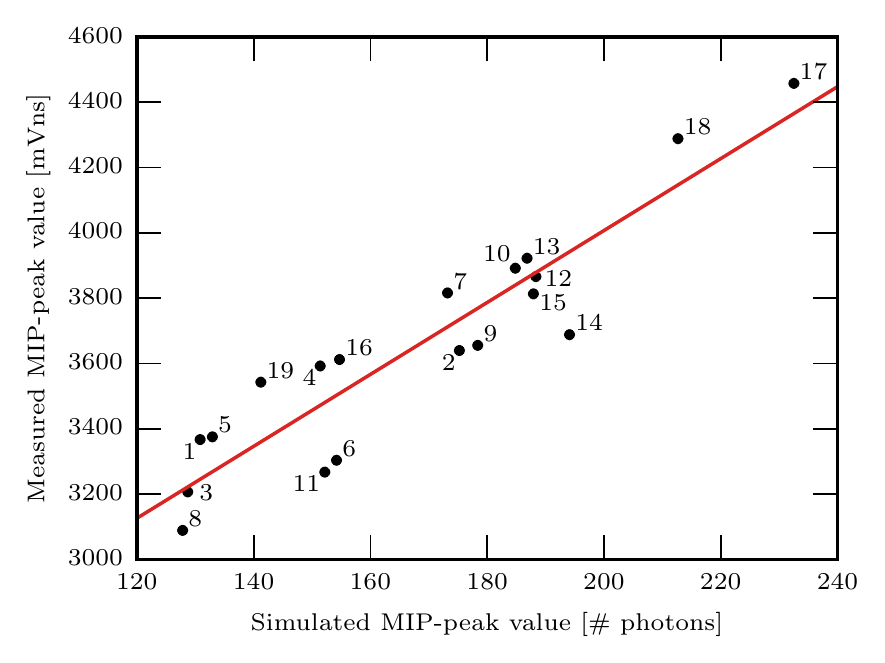}
\caption{Comparison between measured and simulated light yield of a HiSPARC detector (scintillator and light-guide). The numbers in the plot correspond to the locations indicated in Figure \ref{fig:transmission_experiment}.}
\label{fig:transmission_experiment2}
\end{figure}

Figure \ref{fig:transmission_experiment2} shows the comparison between the light-output of the table-top experiment and detector simulation (scintillator and light-guide). Both experimentally measured and simulated MIP-peak values have comparable statistical uncertainties. The simulation reproduces the experimental data rather well. A more detailed analysis including photon arrival times is presented in a separate paper \cite{kasper2019}.

\subsection{Light-guide}
\begin{figure}
\centering
\includegraphics[width=90mm]{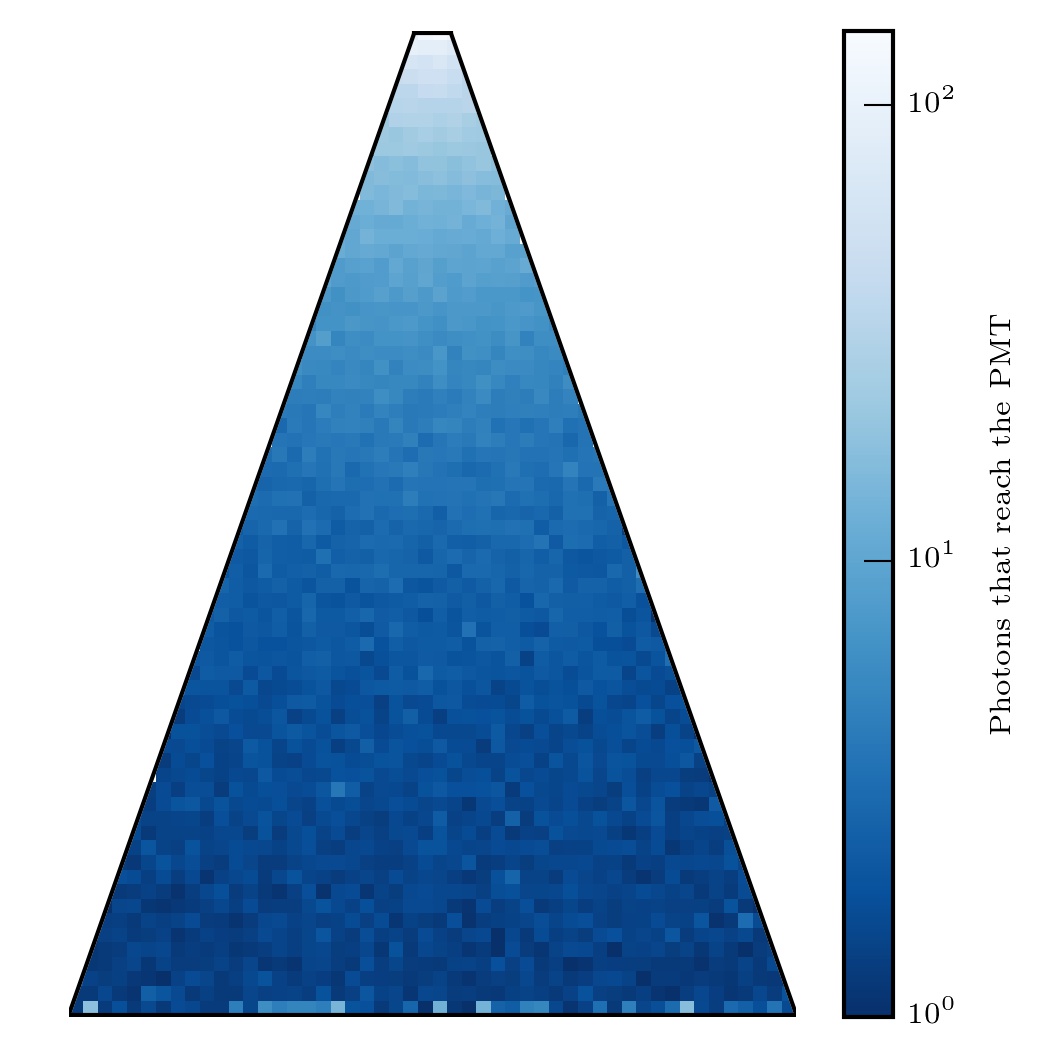}
\caption{Cherenkov light yield as function of position in the light-guide for photons reaching the PMT. In the GEANT4 simulation the light-guide was exposed to perpendicular incident relativistic muons. The muon flux was kept constant over the full area. The probability of measuring Cherenkov photons strongly decreases with increasing distance to the PMT.}
\label{fig:cherenkov}
\end{figure}

The light-guide reduces the scintillator light yield as demonstrated in Figure \ref{fig:sensitivity}. However, the light-guide also may add Cherenkov photons when a charged particle penetrates. Using GEANT4 the production and propagation of these Cherenkov photons have been investigated. The light-guide was exposed to perpendicular incident relativistic muons. Again, the muon flux was kept uniform over the full surface of the light-guide. Figure \ref{fig:cherenkov} shows the spatial distribution at which the Cherenkov photons are released scaled with the number of photons that reach the PMT. The further away from the PMT the Cherenkov photons are generated, the smaller the probability they reach the PMT. Even for large Cherenkov photon yields close to the PMT, the number of photons is small compared to the number of photons generated in the scintillator.

\begin{figure}
\centering
\includegraphics[width=90mm]{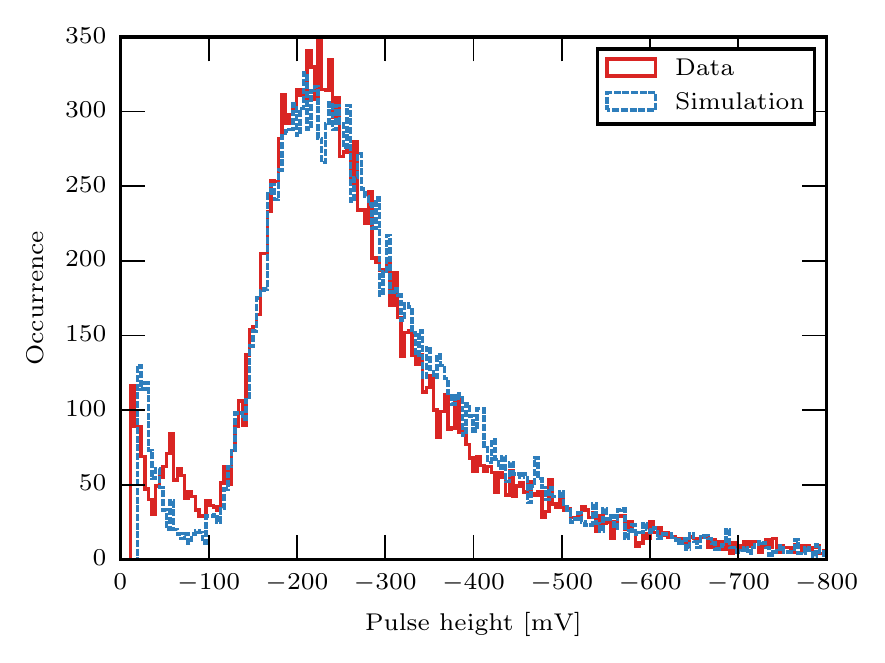}
\caption{Comparison between single muon pulse height distributions of simulated and experimental data. Below $-100$ mV the simulation represents the experimental data rather well. The small peak at $\sim$$-50$ mV is due to a small number of randomly coinciding background noise pulses. The contribution below $-50$ mV is dominated by Cherenkov radiation and is slightly overestimated in the simulation.}
\label{fig:singlemuons}
\end{figure}

Figures \ref{fig:transmission_peaks} and \ref{fig:transmission_experiment2} show the measured signal yield and transmission at selected grid points in the scintillator. To obtain the muon response over the full surface of the detector (scintillator and light-guide) taking into account the proper energy spectrum and angle of incidence of the muon, a second experiment was conducted. Two stacks of each two detectors are placed parallel at a distance of 6 m. When three or more detectors generate a signal exceeding the noise cut-off at $-15$ mV (Fig. \ref{fig:traceexmaple}), the event is recorded. In the analysis, events containing signals with a time difference between both stacks of less than 300 ns are discarded. This excludes the contribution from EASs in which particles arrive in a relatively small time window. If the time difference is larger than 300 ns, the event most likely stems from a random coincidence. These random coincidences are due to single muon background and "noise pulses" (PMT dark pulses, low energy electrons, gamma rays, etc.). When two detectors in the same stack generate a signal exceeding $-15$ mV, a MIP traversed both detectors. If only one detector of a stack generates a pulse, it is most likely caused by 'background noise'. This noise generates a pulse-spectrum that peaks at ($\sim$-50 mV \cite{kasper2019}. The high rate of noise pulses in detectors of the same stack causes only a small number of random background coincidences. By choosing those events in which two detectors in one stack are triggered by a MIP (time difference with a signal in the other stack is larger than 300 ns), single muons are selected. The result is shown in Figure \ref{fig:singlemuons} (red histogram). All four detectors generate a similar pulse height spectrum.

Next, the single muon response in a detector is investigated using the simulation procedure described before. In addition to the propagation of the photons in the scintillator and the light-guide an analytical description of a single photo-electron PMT response is added \cite{dayabay2012}. Muon direction and energy follow the distributions presented in \cite{reyna2006}:
\begin{equation}
I(p,\theta) = \cos^3(\theta)I_\textup{V}(p \cos\theta)
\end{equation}
with $p$ the muon momentum and $\theta$ the zenith angle. The expression connects the relative flux per unit zenith angle to the perpendicular incident muon flux $I_\textup{V}(p)$ [\SI{}{cm^{-2}} \SI{}{sr^{-1}} \SI{}{s^{-1}} \SI{}{GeV^{-1}}]:
\begin{equation}
I_\textup{V}(p) = c_1 p^{-(c_2 + c_3 \log(p) + c_4 \log^2(p) + c_5 \log^3(p))}
\end{equation}
The parameters are $c_1 = 0.00253$, $c_2= 0.2455$, $c_3 = 1.288$, $c_4 = -0.2555$ and $c_5 = 0.0209$ ($\log \equiv \log_{10}$). The energy and zenith-angle are sampled using the Metropolis-Hastings algorithm \cite{hastings1970}. The muons are uniformly distributed across the detector surface. Below $-100$ mV, the simulation (blue histogram) in Figure \ref{fig:singlemuons} compares well with the experimental data. The small peak at $\sim$$-50$ mV is due to the small number of randomly coinciding background pulses in two detectors of the same stack. The small pulse heights stem from Cherenkov radiation in the light-guide. The simulation tends to slightly overestimate the contribution from the Cherenkov photons.

\subsection{Detection efficiency}
The detector simulation describes the experimental data rather well and can be used to investigate the detection efficiency for electrons, muons and gamma rays. The efficiency depends not only on a combination of gain (i.e. precise value of the MIP-peak) and applied signal threshold (defined in terms of the fraction of the MIP-peak value), but also on the energy of the particle and its angle of incidence. A lower limit on the efficiency is obtained when only perpendicular incident particles are considered. The detection efficiency is then defined as the fraction of the pulse height distribution exceeding the threshold divided by the full range of pulse heights. Only the area covered by the scintillator is considered; the contribution from Cherenkov photons in the light-guide is ignored.

\begin{figure}
\centering
\includegraphics[width=90mm]{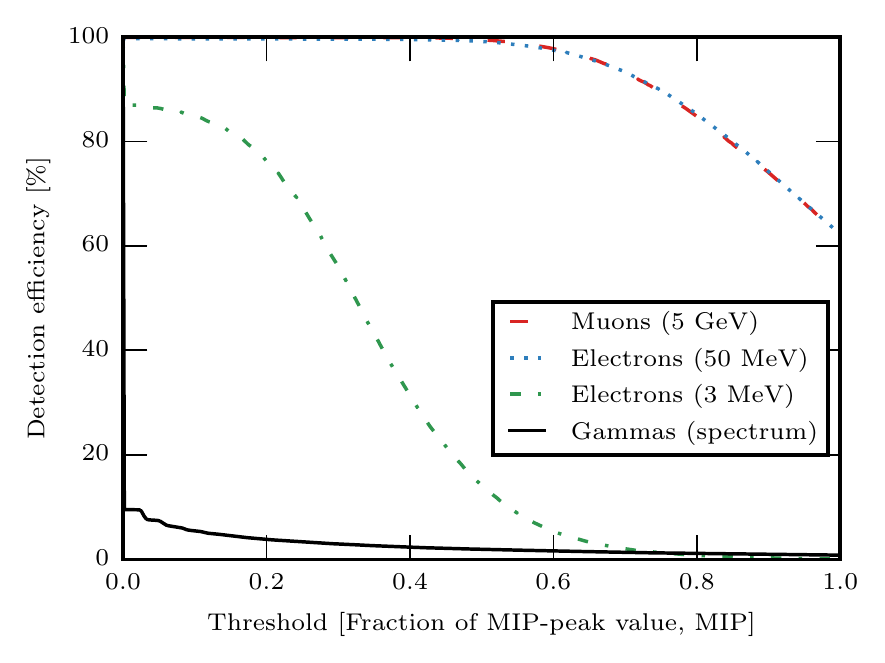}
\caption{Detection efficiency of a detector (only the area covered by the scintillator is considered) as a function of threshold (fraction of the MIP-peak value) for muons, electrons and gamma rays. The MIP-peak is defined as the most probable signal response of a single minimum ionizing particle. For high energy electrons (50 MeV, blue, dotted line) and energetic muons (5 GeV, red dashed line) the efficiency curves coincide. For low energy electrons (3 MeV, green dash-dotted line) the efficiency drops significantly. For gamma rays, the full energy range is considered. The energy spectrum is obtained from air shower simulations. The gamma detection efficiency turns out to be small (black solid line).}
\label{fig:trigger_efficiency}
\end{figure}

Figure \ref{fig:trigger_efficiency} shows the detection efficiency curves for the various particles. The majority of muons is considered to be minimum ionizing as they are produced at high altitudes. Electrons however, are also generated in electromagnetic showers closer to ground. Low energy electrons are therefore more abundant than low energy muons. For high energy muons (5 GeV) and high energy electrons (50 MeV) the detection efficiencies are the same (dashed red and dotted blue lines coincide). This is to be expected since at these energies both particles are minimum ionizing. For thresholds up to 0.5 MIP their detection efficiency is very close to 100\%. For low energy electrons (e.g. 3 MeV), the detection efficiency is very different (green dashed-dotted line) and stays well below 100\% over the full range of threshold values. Low energy particles have a high chance to be absorbed in the scintillator and will therefore produce only a limited number of excitation photons.

Gamma rays are abundant in EASs. Their energy spectrum is obtained from EAS simulations (CORSIKA \cite{corsika1998}). The majority of gamma rays will not interact in the scintillator; their detection efficiency will therefore be small. For perpendicular incident gamma rays, a lower limit on their detection efficiency as function of threshold is also depicted in Figure \ref{fig:trigger_efficiency} (solid black line). Only a fraction (less than  10\%) of the gamma rays will be detected.

\section{HiSPARC station}
\label{sec:hisparcstation}
A HiSPARC station combines two or four detectors with the aim to distinguish EASs from single background muons. Since the arrival times of particles in an EAS are highly correlated, this is achieved by demanding a response in two or more detectors within a small time frame.

\begin{figure}
\centering
\includegraphics[width=90 mm]{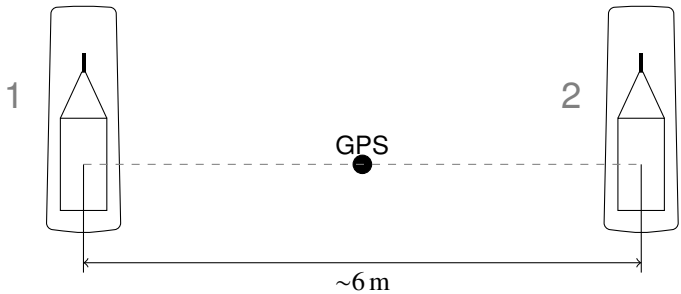}
\caption{Typical configuration of a two-detector station. A GPS antenna is located in between the two detectors and provides a signal to precisely timestamp the arrival time of the EAS particles.}
\label{fig:stationlayout2}
\end{figure}

\begin{figure}
\centering
\includegraphics[width=90 mm]{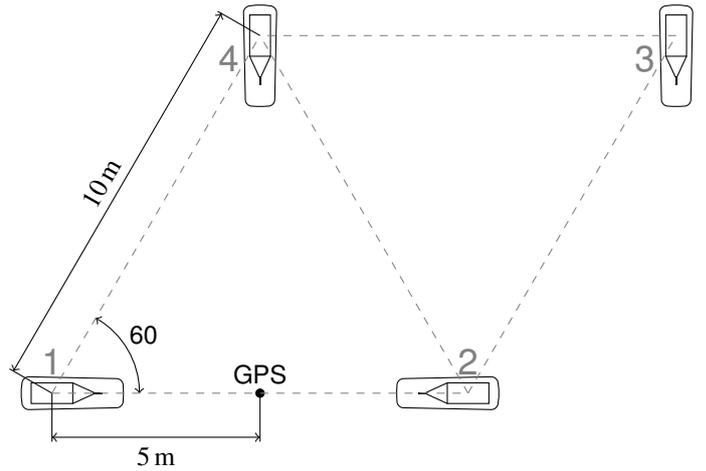}
\caption{Four detectors placed in a diamond formation.}
\label{fig:stationlayout4a}
\end{figure}

\begin{figure}
\centering
\includegraphics[width=90 mm]{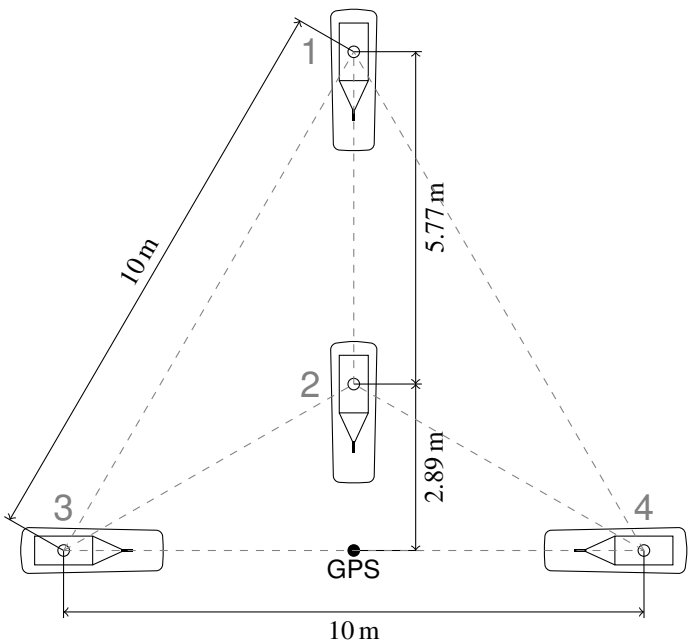}
\caption{Four detectors placed in a triangle formation with the fourth detector at the centroid.}
\label{fig:stationlayout4b}
\end{figure}

The layout of a HiSPARC station with two detectors is shown in Figure \ref{fig:stationlayout2}. The majority of high schools deploy a two-detector station. Four-detector stations explore two different layouts; a diamond formation (Fig. \ref{fig:stationlayout4a}) and an equilateral triangle with one detector at the centroid of the triangle (Fig. \ref{fig:stationlayout4b}). When at least three detectors in a four-detector station observe one or more particles of an EAS, the direction of the EAS (and thus the direction of the primary cosmic ray) can be obtained by triangulation. When only two detectors are hit, as for a two-detector station, the time difference between the two detectors only allows for the reconstruction of the arrival direction along the axis that connects the two detector centers.

Changes in atmospheric pressure are of importance; the higher the pressure, the larger the probability for a (low energy) shower particle to be absorbed before reaching the ground. Consequently, this will affect the size of the footprint. The pressure also influences at which height the first interaction occurs. Currently, 19 HiSPARC stations are equipped with a weather station \cite{DavisVantagePro}. Weather data are collected together with the cosmic ray data and stored in the Nikhef central database.

\subsection{Read-out electronics}
\label{sec:electronics}
\begin{figure}
\centering
\includegraphics[width=90 mm]{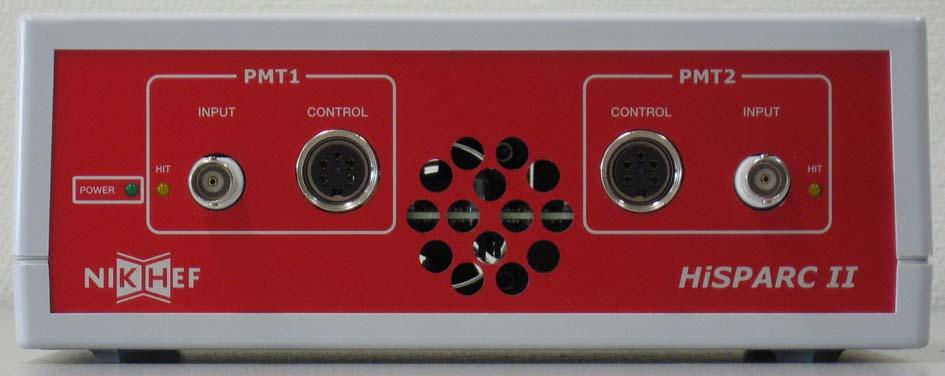}
\includegraphics[width=90 mm]{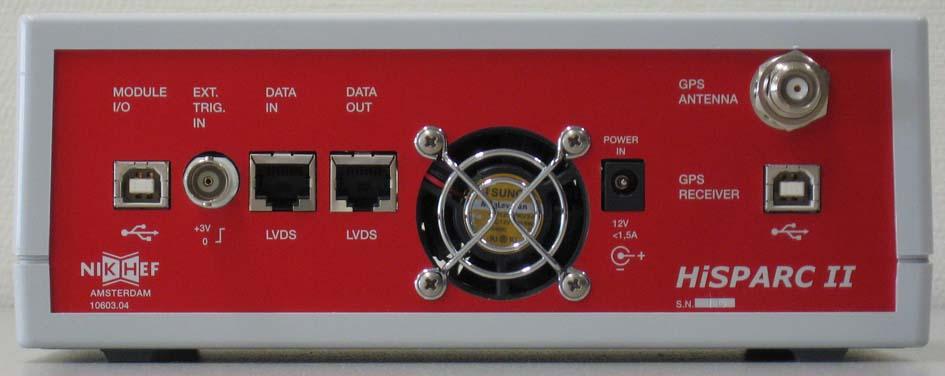}
\caption{Front (top) and back (bottom) of the HiSPARC readout unit. Two PMTs are connected to the front of the unit and supplied with $+12$ V DC and a reference voltage. In the PMT-base, this reference voltage ($+0.3$ to $+1.5$ V) is converted into  a high voltage ($-300$ to $-1500$ V). An orange LED next to the signal input flashes when the input signal ($-10.3$ to $+0.3$ V) exceeds the lower threshold. A white LED behind the air outlet (center) flashes when a trigger condition is met. At the back of the unit there are connectors for the 1.5 A 12 V DC power supply, GPS antenna cable and USB port for monitor and control of the GPS unit. At the far left there is a USB connector for data output. Two UTP ports facilitate the communication between Master and Slave units. There is also an input for an ($+3$ V) external trigger. The ADCs are set to a dynamical range of $-2.2$ V to $+0.1$ V. The maximum trigger rate is in excess of 30 events per second.}
\label{fig:hisparcunit}
\end{figure}

Detectors are connected to read-out electronics by cables with a standard length of 30 m. The custom designed electronics control and read out two PMTs (Fig. \ref{fig:hisparcunit}). One electronics unit facilitates two detectors. For a four-detector station two units are connected in Master-Slave configuration. All four detectors are treated exactly the same. Their signals can be used to construct a matrix of trigger conditions.

The digitisation of the analog input signal is carried out by two 12 bits, 200 MHz analog-to-digital converters (ADCs). One ADC is triggered at the rising edge of the clock, the other ADC at the falling edge, thus doubling the sampling frequency to 400 MHz, i.e. at 2.5 ns intervals. A calibration procedure (ADC 'alignment') ensures that both ADCs yield the same baseline, gain and dynamical range ($\sim$$+0.1$ V to $\sim$$-2.2$ V or 0 - $\sim$15 MIPs). For larger signals two comparators with adjustable threshold (default at $-2.5$ V ($\sim$17 MIPs) and $-3$ V ($\sim$20 MIPs) resp.) are added. The comparator data significantly improve the off\-line reconstruction of large signals that exceed the dynamical range of the ADCs \cite{TimKokkeler}.

\begin{figure}
\centering
\includegraphics[width=90 mm]{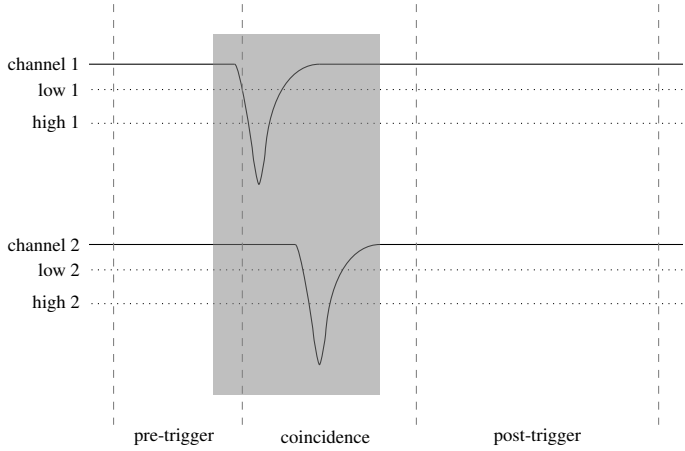}
\caption{A schematic representation of an event. Dashed vertical lines: the pre-trigger (\SI{1}{\micro\second}), coincidence (\SI{1.5}{\micro\second}) and post-trigger (\SI{3.5}{\micro\second}) windows. Shaded area: the data-reduction window. Data outside this window will not be stored. The dotted lines indicate the low and high threshold values.}
\label{fig:triggerexplanation}
\end{figure}

The output of the ADCs is transferred into embedded memory in a field-programmable gate array (FPGA). The FPGA is clocked at 200 MHz as well. Per channel two thresholds can be defined: a low threshold and a high threshold. The FPGA raises flags when a signal exceeds a thres\-hold. Combining flags (AND/OR) for up to four detector channels provides an extensive matrix of trigger conditions. For test purposes the unit has a separate external trigger input that can be combined with this trigger matrix. Figure \ref{fig:triggerexplanation} gives an example on how a trigger condition for a two-detector station is generated and how an event is composed.

Assume the trigger condition is defined such that both detectors have to generate a signal that exceeds the low threshold within a limited amount of time. The time window has to be large enough to be fully efficient for detecting particles belonging to the same EAS (with different arrival times, covering all possible angles of incidence of the shower), while it has to be small enough to minimize the random number of coincidences between background muons. In the figure a first flag is raised when detector 1 generates a signal that exceeds the low threshold (channel 1) - the signal may also be large enough to raise the flag for exceeding the high threshold -. At the same time a 'coincidence' time window is opened. The length of this time window is typically of order \SI{1.5}{\micro\second}. If during the \SI{1.5}{\micro\second} the second detector also generates a signal that exceeds the low threshold, the trigger condition is met and an event is generated. An event consists of data taken just before the coincidence window was opened (the pre-trigger time window, typical length \SI{1}{\micro\second}), the coincidence time frame (\SI{1.5}{\micro\second}) and post-trigger period; the post-trigger time window (\SI{3.5}{\micro\second}).

The maximum length of an event window is \SI{10}{\micro\second} (= 2000 12 bit ADC samples = 3kB memory). Per detector channel, the maximum event size becomes 6 kB. The (embedded) memory may contain multiple events (maximum 3 for the time windows defined above) and acts as a de-randomizing buffer. Events are transferred from the FPGA's embedded memory into a USB buffer from where they are read by the DAQ PC. If the internal memory is full and the USB buffer is not read out fast enough, new events are discarded until an event is transferred to the DAQ PC. In practice, the trigger rate remains below 1 Hz for all three station layouts. The probability that a fourth event immediately occurs after three consequent triggers (\SI{18}{\micro\second}) is negligible.

A small GPS module \cite{gpsmodule} is mounted on top of the electronics mother board. It generates one second 'tick marks' that are sent to the FPGA that generates a separate message for the DAQ software. Each second a counter is started that counts until the trigger flag is raised or the next one second information arrives. In addition to the one second information (in ns) and value of the counter, the message also includes the number of GPS satellites and their signal strength. The DAQ software combines three one second messages, the counter values for the intervals and GPS quantisation error to calculate the precise time stamp of an event.
In a four-detector station a 'Slave unit' is connected to the 'Master'. Both Master and Slave contain the same electronics and FPGA firmware. However, the Slave unit has the GPS module removed. By connecting the two units (Fig. \ref{fig:hisparcunit}), the trigger matrix is extended to four channels. The individual Master and Slave messages are combined by the DAQ software to produce the full event. Each second and for all channels, also the number of times a signal exceeds the high and/or low threshold is recorded.
 
There exist two versions of the electronics. They have the same functionality. The HiSPARC II unit has an on-board memory chip that contains firmware that is loaded into the FPGA at power-up. The HiSPARC III box has an additional USB channel via which the DAQ software transfers the firmware to the FPGA at run-time.

\subsection{DAQ software}
\begin{figure*}
\centering
\includegraphics[width=140mm]{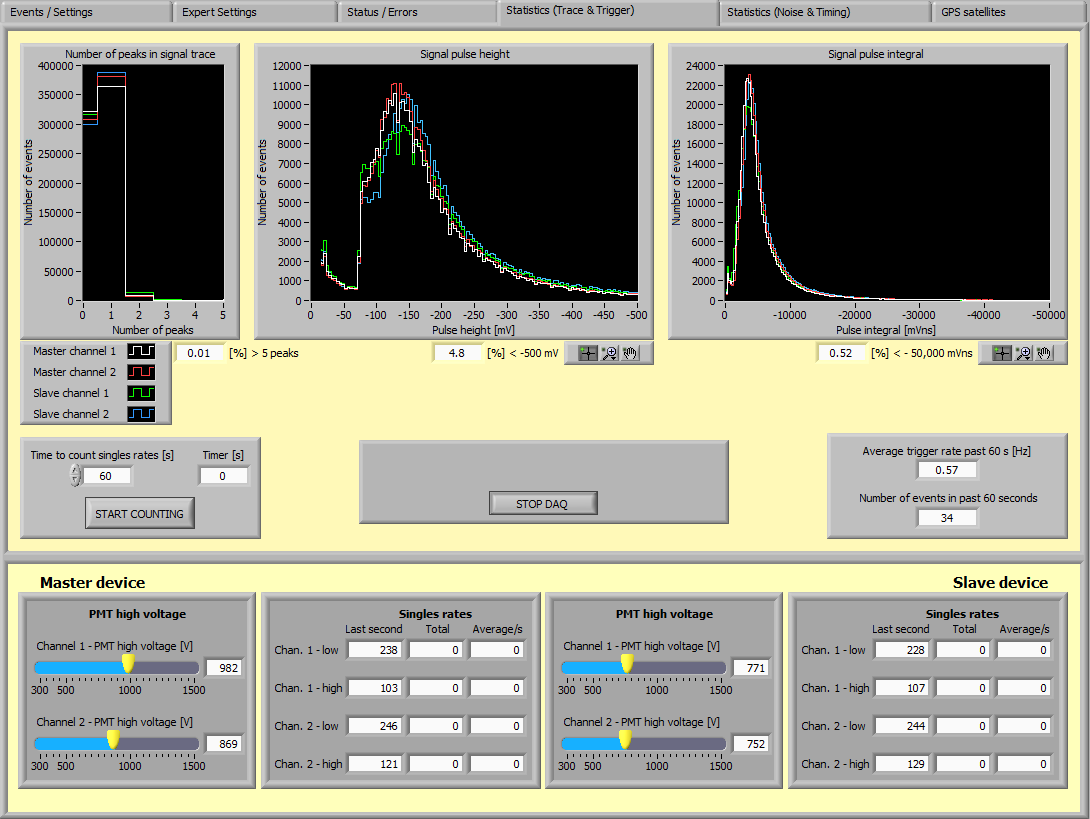}
\caption{Screenshot of a panel in the HiSPARC DAQ control and monitor user interface (colour online). The top three histograms show (from left to right) the number of signals above threshold in an event for each detector, the pulse heights and the pulse integral distributions resp. The bottom control panels set the PMT high voltage for Master (left) and Slave (2\textsuperscript{nd} from right). The other panels display single rates for signals exceeding low and high thresholds. Additional tabs (visible at the top of the screenshot) include event settings, status and error messages, and display GPS data.}
\label{fig:daq}
\end{figure*}

To control the electronics unit and to monitor the data, a graphical user interface based on LabVIEW\texttrademark  \cite{labview2007} has been developed that executes on a Windows\texttrademark  PC. A screenshot of one of the interface panels is shown in Figure \ref{fig:daq}. Typical pulse height distributions for a four-detector station are shown in the central graph. On the right the pulse integral distributions are displayed. With a typical pulse length of 30 to 50 ns (Fig. \ref{fig:traceexmaple}) a zero-suppression algorithm is applied to reduce event sizes.

The DAQ software performs a preliminary analysis of the data. The baseline (which is also used by the zero suppression algorithm) and the fluctuation of the baseline are recorded. Subsequently, the pulse height and pulse integral are calculated. The software also keeps track of the number of accessible GPS satellites and their signal strength over time.

The DAQ software stores the raw data from the electronics unit(s) combined with the first analysis results into a local MySQL \cite{mysql} database. A Python \cite{python} program monitors the number of written events and uploads the data at regular intervals to the data-store server at Nikhef.

The Nagios \cite{nagios} software package monitors the status of both hardware (electronics and PC) and DAQ software for each station. In case a service fails an e-mail is generated and automatically forwarded to the person responsible for the station with the request to intervene. To avoid down time during weekends and long high school (summer) breaks, Nikhef operates an OpenVPN \cite{openvpn} network that gives remote access to each station (TightVNC \cite{tightvnc}). This network is also used to install software updates on the stations.

\subsection{Trigger efficiency}
The default trigger condition for a two-detector station is straightforward. An event is selected if the signals in the two detectors exceed the low threshold within the coincidence time window (\SI{1.5}{\micro\second}). The default trigger conditions for a four-detector station are: the signals of at least two detectors pass the high threshold or at least three detectors generate a signal that exceeds the low threshold.

Figure \ref{fig:trigger_efficiency} demonstrated that the detection efficiency for perpendicular incident MIPs traversing a detector is $T_\textup{L}=0.999$ (0.2 MIP) for the low threshold and $T_\textup{H}=0.997$ for the high threshold ($\sim$0.5 MIP). For a two-detector station this leads to a trigger efficiency of $T_\textup{L}^2 > 0.99$. For a four-detector station the trigger efficiency depends on the number of detectors that are hit. For two MIP events the trigger probability depends on the high threshold: $T_\textup{H}^2 > 0.99$. For three MIP events, for which it is unknown which three detectors are hit, the station will still trigger even if one of the particles is not detected ('1 non'). If one of the MIPs does generate a signal that goes over the low threshold, the two other MIPs may still generate a signal that goes over the high threshold (2 high). This combination occurs three times:
\begin{align}
\begin{split}
P^{\textup{four}}_\textup{3mip} &= P(\textup{3 low}) \\
&\quad+ 3 \cdot P(\textup{2 high}|\textup{1 non})P(\textup{1 non}) \\
&= T_\textup{L}^3 + 3 \cdot T_\textup{H}^2 \cdot (1-T_\textup{L}) \\
&\gg 0.99
\end{split}
\end{align}
Finally, for four MIPs crossing four different detectors the station is also very efficient:
\begin{align}
\begin{split}
P^{\textup{four}}_\textup{4mip} &= P(\textup{4 low}) \\
&\quad + 4 \cdot P(\textup{3 low}|\textup{1 non})P(\textup{1 non}) \\
&\quad+ 6 \cdot P(\textup{2 high}|\textup{2 non})P(\textup{2 non}) \\
&= T_\textup{L}^4 + 4 \cdot T_\textup{L}^3 \cdot (1-T_\textup{L}) \\
&\quad+ 6 \cdot T_\textup{H}^2 \cdot (1-T_\textup{L})^2 \\
&\gg 0.99
\end{split}
\end{align}
In all cases the trigger efficiency of HiSPARC stations for MIPs is well above 99\%. In practice not only MIPs will be detected. Also gamma rays, low energy electrons and low energy muons are part of the EAS. The detection efficiency then strongly depends on their energy and nature.

\subsection{Timing offsets}
\begin{figure}
\centering
\includegraphics[width=90mm]{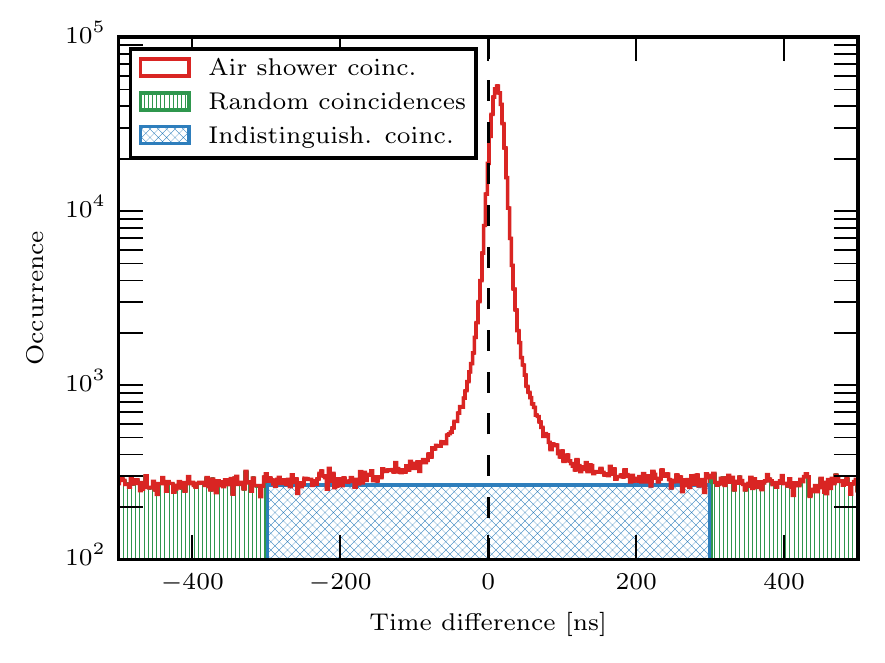}
\caption{Timing offset between two detectors in a two-detector station (in red). The plateau (blue horizontal line) is due to random coincidences (uncorrelated particles, PMT noise etc.). For time differences smaller than 300 ns (blue crossed region) the random coincidences are indistinguishable from air shower coincidences. Their contribution is however small.}
\label{fig:timedifferences}
\end{figure}

Assuming that the distribution of the angle of incidence of EASs is isotropic, each detector of a station has about equal probability to be hit first. Figure \ref{fig:timedifferences} shows the (arrival) time differences between two detectors in a two-detector station (station \#4). About $8 \times 10^5$ events were collected between the 5\textsuperscript{th} of March and the 2\textsuperscript{nd} of April 2018. The shift of the peak with respect to the dashed line at 0 ns shows an average timing offset between the two detectors of 12 ns. A plateau (blue horizontal line) is reached for offsets larger than $\sim$300 ns. This is the result of random coincidences, i.e. two uncorrelated particles, a particle and a spontaneous emission in the PMT etc. The height of the plateau is obtained by making a fit for offsets between 300 ns and \SI{1.5}{\micro\second} (default coincidence time window). The expected number of random coincidences per second ($N$) can be calculated with $N = 2 \tau r_1 r_2$. $\tau$ is the coincidence time window and $r_1$ and $r_2$ are the recorded single rates in the detectors. Integration over the full length of the plateau region yields $3.21 \times 10^5$ events. This is well in agreement with the estimated number of random coincidences: $3.17 \times 10^5$. Note that the number of random coincidences within a time difference of 300 ns (blue crossed region) is small.

\begin{figure}
\centering
\includegraphics[width=90mm]{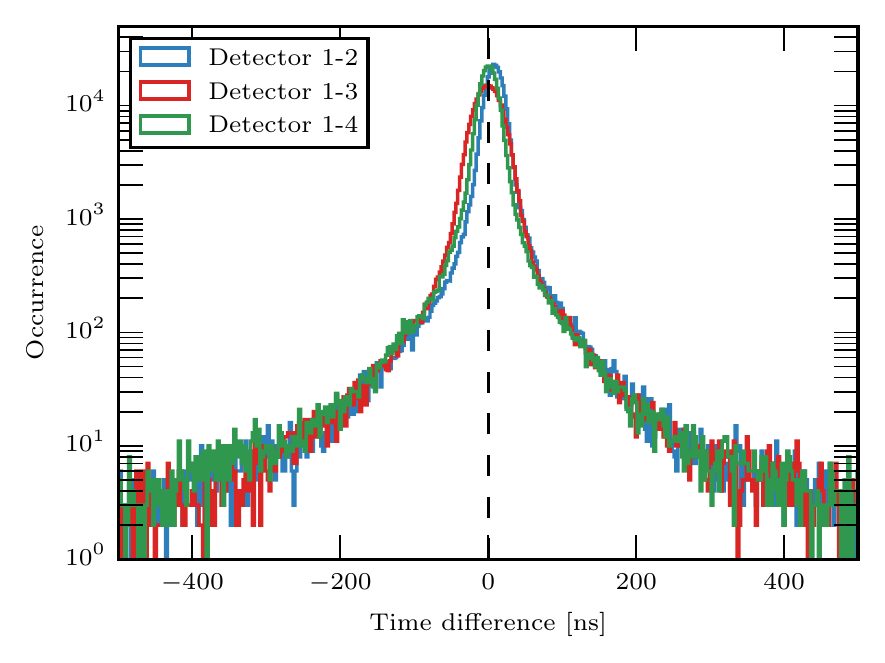}
\caption{Timing offsets between detectors in a four-detector station (colour online). Only events in which all detectors generated a signal exceeding the low threshold are taken into account. There is no plateau due to the small probability that a random coincidence between the four detectors occur at the same time. The distance between detectors 1 and 2, and detectors 1 and 4, are equal and show an almost identical time offset distribution (in blue and green resp.). The distance between detectors 1 and 3 is larger; they lie along the diagonal of the diamond. The time difference distribution (in red) is therefore slightly broader.}
\label{fig:timedifferencesfour}
\end{figure}

The difference in arrival times in detectors of a four-detector station are shown in Figure \ref{fig:timedifferencesfour}. Station \#501 (diamond formation, Fig. \ref{fig:stationlayout4a}), recorded $3.1 \times 10^5$ events from the 1\textsuperscript{st} of January till the 1\textsuperscript{st} of March 2018. Only time differences in events for which all four detectors generated a signal exceeding the low threshold are considered. The random coincidence plateau is absent. The probability that four random signals coincide within \SI{1.5}{\micro\second} is thus very small. The timing offset in detector combinations 1-2 (blue) and 1-4 (green) show almost identical distributions. The distance between the detectors in the two pairs is the same. The distance between detectors 1 and 3 is larger as they lie along the diagonal of the diamond. The time offset distribution for this combination (in red) is therefore slightly broader.

For each station the timing offsets are calculated and stored on a daily basis. They are important parameters in for instance the direction reconstruction of an EAS.

\section{Clusters of stations}
\label{sec:clusterperformance}
\begin{figure}
\centering
\includegraphics[width=90mm]{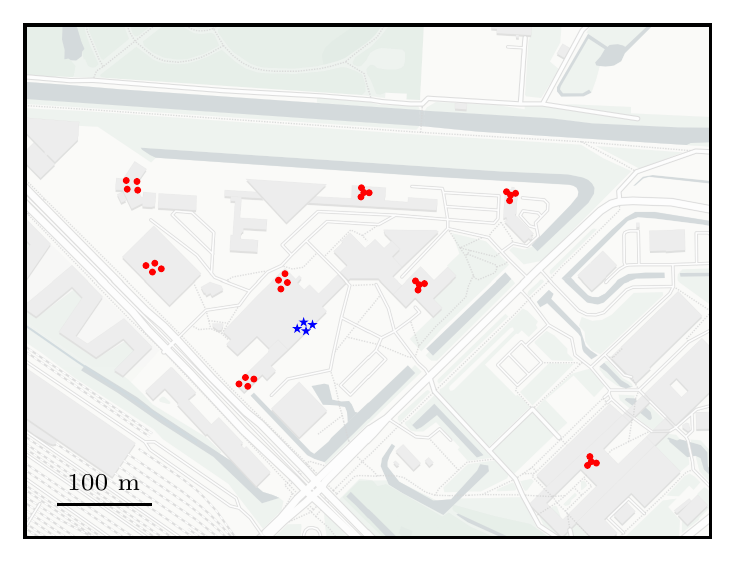}
\caption{Location of stations at the Amsterdam Science Park cluster. One of the stations is located inside the Nikhef building and is not on the map. Each red dot represents a detector and each combination of four dots forms a station. There are diamond and triangle shaped station configurations. At one location, four 'diamond' stations are essentially positioned 'on top of each other' (detectors are $\sim$1 m apart) and only a single station is displayed (blue stars).}
\label{fig:sciencepark}
\end{figure}

A HiSPARC cluster is a collection of stations. The surface that the cluster covers often depends on the number of stations that a high school hosts and/or how many high schools there are in the neighborhood. For extensive research a dense cluster at the Amsterdam Science Park is created (Fig. \ref{fig:sciencepark}). The Science Park cluster contains thirteen four-detector stations. The distance between the stations varies from $\sim$1 m up to 280 m. Several stations may sample the same (large) EAS while each station generates its own GPS time stamp. To reconstruct the angle of incidence of the primary cosmic ray the arrival time of EAS particles in each station has to be precisely known. This implies that it is crucial to account for time offsets between GPSs.

\subsection{GPS timing}
Each GPS is operated in 'overdetermined clock mode' and is at a fixed location. Upon installation, the GPS receiver performs a 24 hour self-survey to accurately determine its position which, in turn, also provides the absolute coordinates for the detectors of that station. The precision of the self-survey is investigated by performing multiple self-surveys using various combinations of GPS antennas, cables and GPS modules (of same type and make). These systematic studies show that 50\% of the longitudinal differences stay within 0.73 m, 75\% within 1.8 m and 95\% within 4.1 m while for the latitude differences are 50\% within 0.48 m, 75\% within 1.1 m and 95\% within 2.6 m. Analysis of the altitude data shows that 50\% of the combinations stay within 1.5 m, 75\% within 2.8 m and 95\% within 6.1 m. The manufacturer of the GPS electronics quotes a $1\,\sigma$ accuracy of 15 ns \cite{gpsmodule}. When combining data from multiple stations in reconstructing an EAS, the precise timing offsets between all stations in a cluster have to be known. By replacing the detectors by stations, the same method can be applied with which the timing offsets between detectors within one station were obtained. Again, the angle of incidence of EASs is assumed to be isotropically distributed; stations have equal probability to be hit first. The timing offsets between $\sim$100 station pairs have been examined. Combining their offsets, a Gaussian distribution with $\mu = 2.7$ ns and $\sigma = 18.9$ ns is obtained. As the distance between the majority of the stations is (much) larger than the distance between the detectors within a station, the rate at which coincidences between stations occur is much smaller than the event rate in a station. Consequently, GPS offsets can usually not be derived on a day-to-day basis due to lack of statistics.

\subsection{Acceptance}
Obviously, a station near the equator covers a different part of the sky when compared to a station near, for instance, the North Pole. Moreover, with increasing angle of inclination w.r.t. the zenith, the distance the shower particles have to travel through the atmosphere increases. This implies that for the same energy of the primary particle the number of shower particles that reach the ground decreases with increasing zenith angle.

The zenith angle ($\theta$) dependent rate distribution integrated over all energies can be approximated by: $dN/d\theta \propto 2\pi \sin \theta \cos^\beta \theta$ with $\beta = 6$ \cite{wilson1956}. The rate for small zenith angles is therefore suppressed ($\theta \to 0^{\mathrm{o}}$). The zenith angle dependent flux is obtained by dividing by the geometrical factor $2 \pi \sin\theta$ (thus a high flux for small zenith angles). Next, the coordinate system in which the flux (integrated over azimuth) is defined, is converted into an equatorial coordinate system with $\alpha$ the right ascension and $\delta$ the declination. In addition, the longitude, latitude and altitude of the station, and the rotation of the Earth are taken into account. The acceptance is calculated in this coordinate system while the flux is integrated over 24 hours.

\begin{figure}
\centering
\includegraphics[width=90 mm]{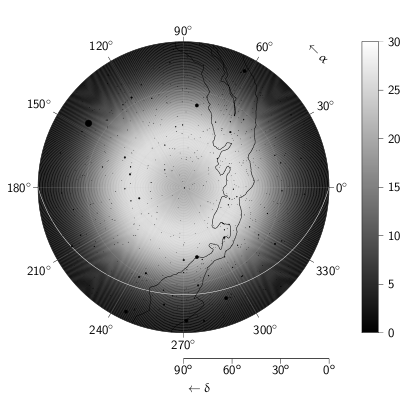}
\caption{The acceptance of an Amsterdam Science Park station integrated over a day. The skymap is represented in equatorial coordinates. Due to the Earth's rotation cosmic ray sources will only be visible part of the day. The integrated acceptance over a full day is shown in grey-scales from 0\% (black) to a maximum of 30\% (white). The boundaries of the Milky Way are shown as black curved lines while major stars are indicated by black dots. The ecliptic is shown in white. Note that only cosmic rays from the Northern Hemisphere can be observed.}
\label{fig:acceptance}
\end{figure}

We assume that an EAS generated by a source directly above Amsterdam Science Park ($\theta= 0^{\mathrm{o}}$) has an acceptance of 100\%. As the Earth rotates the zenith angle increases and the number of EAS particles that reach the ground decreases. At some point the source disappears behind the horizon. The overall acceptance integrated over 24 hours becomes $\sim$30\% (light 'donut' shaped band in Fig. \ref{fig:acceptance}). Sources at $\delta=0^{\mathrm{o}}$ will be below the horizon most of the time. As a result, the acceptance for primaries stemming from the equatorial plane diminishes \cite{NiekArxiv}. The maximum acceptance is obtained at equatorial directions that match small zenith angles in the local coordinate system of the station. Despite the fact that the area around Polaris is always visible from Amsterdam Science Park, and that the acceptance is relatively high, it never reaches 100\% since the zenith angle is always larger than zero.

\section{Data processing}
\label{sec:dataprocessing}
All stations ($\sim$140) send their data to Nikhef where they are stored in HDF5\textsuperscript{\circledR} \cite{hdf5} format. Each night, a Django \cite{django} web application preprocesses the raw data and generates an event summary database (ESD). The ESD contains information such as the pulse heights and timestamp of an event, detector position and timing offsets, etc. The same application supports direct web access to the ESD. It also provides an application programming interface (API) through which also raw data are publicly accessible.

For data manipulation a Python based module has been developed. The SAPPHiRE (Simulation and Analysis Program Package for HiSPARC Research and Education) \cite{sapphire2012} facilitates downloading HiSPARC data and performing analyses. It contains an extensive set of EAS reconstruction tools. In addition to the Python module a JavaScript library is developed called jSparc \cite{jSparc}. With jSparc students can work with (or create their own) web applications to explore the HiSPARC data.

The SAPPHiRE (Python) library also forms the basis for a set of Jupyter Notebooks \cite{jupyternotebook}. These notebooks are developed for use in the classroom and for (high school) student research projects.

\section{EAS direction reconstruction}
\label{sec:easdirectionreconstruction}
In section \ref{sec:detector} a detailed simulation of the single particle response of a HiSPARC detector was presented. Using this detector simulation, the response of a four-detector HiSPARC station to EASs can be investigated. With CORSIKA \cite{corsika1998}, proton initiated EASs are generated with energies ranging from $10^{13}$ to $10^{16.5}$ eV. Their relative abundance follows the cosmic ray energy spectrum. 'Thinning' \cite{hillas1981} was not applied. For high energy hadronic interactions the QGSJET-II \cite{qgsjet2006} model is selected. Interactions of hadrons with energies below 80 GeV are simulated using GHEISHA \cite{gheisha1985}. Electromagnetic interactions are described by the EGS4 \cite{egs41985} model. While the location of the station in the simulations remains fixed, the position of the EAS core is randomly chosen within a circle with a radius of 100 m centered at the station. Arrival directions are chosen isotropically. When one or more EAS particles hit a detector the full detector simulation is applied. For events satisfying the trigger conditions and having at least two MIPs in each detector, the direction of the shower is reconstructed assuming a flat shower front using the (triangulation) algorithm described in \cite[Chapter~5]{proefschriftHans}.

\begin{figure}
\centering
\includegraphics[width=90mm]{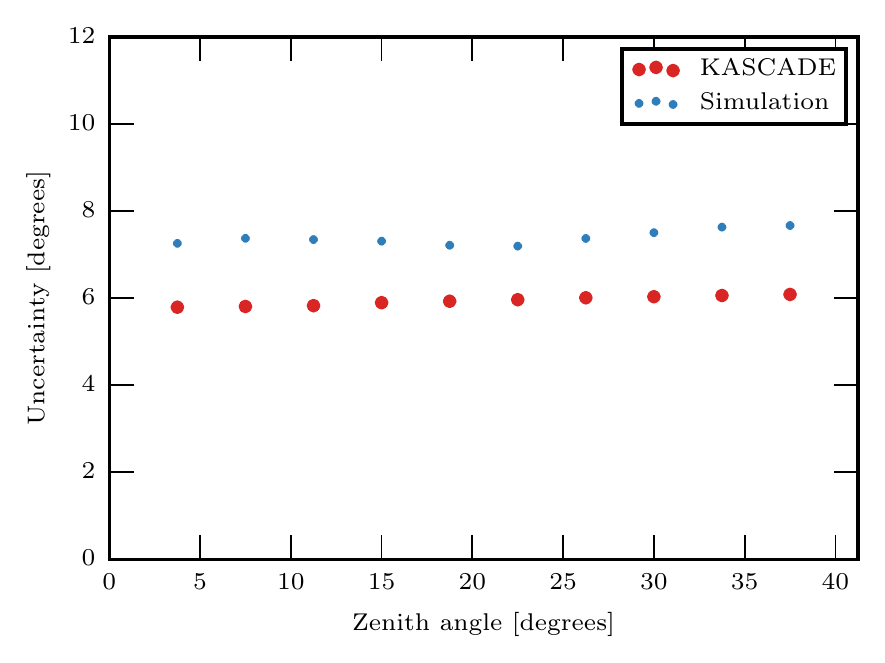}
\caption{The $1\,\sigma$ uncertainty in the shower direction reconstruction as a function of zenith angle (blue dots) is calculated by comparing the direction of CORSIKA generated showers and the direction reconstructed after full detector simulation in a four-detector, triangle shaped, HiSPARC station. A station with the same configuration was integrated in the KASCADE experiment. The reconstruction algorithm that was applied to the simulated EASs was used to obtain the shower directions from the HiSPARC data in the KASCADE setup. The algorithm only used events (satisfying the trigger conditions) with at least 2 MIPs in each detector. The comparison between the shower direction reconstructed by KASCADE ($0.3^{\mathrm{o}}$ accuracy) and measured by the HiSPARC station is shown as a function of zenith angle (red dots).}
\label{fig:directionreconstruction}
\end{figure}

Figure \ref{fig:directionreconstruction} gives the $1\,\sigma$ uncertainty in the shower direction reconstruction for a four-detector triangle shaped station (Fig. \ref{fig:stationlayout4b}) as a function of zenith angle (blue dots). The distribution is obtained by comparing the direction of the primary cosmic ray as set in the CORSIKA Monte Carlo program and the reconstructed direction after detector simulation. The average uncertainty ($\theta < 40^{\mathrm{o}}$) is $7.7^{\mathrm{o}}$.

\begin{figure}
\centering
\includegraphics[width=90mm]{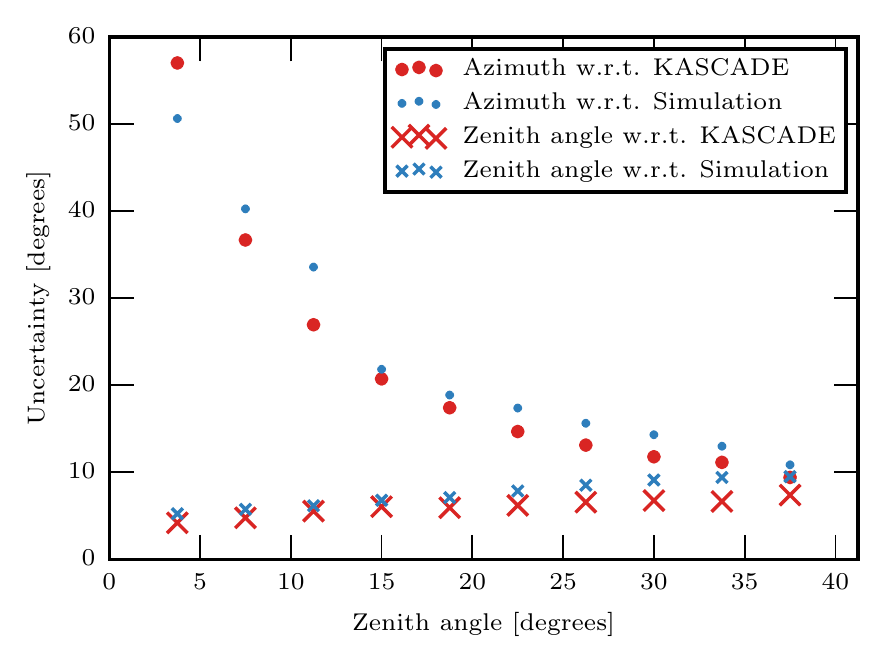}
\caption{The $1\,\sigma$ uncertainty in the reconstruction of azimuth (blue dots) and zenith (blue crosses) as a function of zenith angle are calculated by comparing the direction of CORSIKA generated showers and the direction reconstructed after full detector simulation in a four-detector, triangle shaped, HiSPARC station. A station with the same configuration was integrated in the KASCADE experiment. The reconstruction algorithm that was applied to the simulated EASs was used to obtain the shower directions from the HiSPARC data in the KASCADE setup. The reconstruction algorithm only used events (satisfying the trigger conditions) with at least 2 MIPs in each detector. The comparison between the shower direction reconstructed by KASCADE ($0.3^{\mathrm{o}}$ accuracy) and measured by the HiSPARC station, is shown for azimuth (red dots) and zenith (red crosses) as a function of the zenith angle.}
\label{fig:reconstructionsimulations}
\end{figure}

The shower direction can be decomposed in terms of zenith angle and azimuth. Figure \ref{fig:reconstructionsimulations} shows the $1\,\sigma$ uncertainty in the azimuthal angle (blue dots) and zenith angle (blue crosses) as a function of zenith angle. Again, the distribution is obtained by comparing the direction in which the shower developed as set in the CORSIKA Monte Carlo program and the reconstructed direction after detector simulation. With increasing zenith angle, the $1\,\sigma$ uncertainty on the reconstructed zenith angle slightly increases. The uncertainty on the reconstructed azimuth however, rapidly increases for smaller zenith angles; in the limit where the zenith angle goes to zero the azimuth becomes undefined.

In 2008 a four-detector HiSPARC station (triangle configuration) was integrated in the KASCADE experiment \cite{kascade2003}. When KASCADE detected an EAS in the area where the HiSPARC station was located, the experiment generated a signal that triggered the DAQ of the HiSPARC station. Between July 1 and August 6 2008 more than $5 \times 10^5$ events were recorded \cite{proefschriftDavid}. The direction of these showers was reconstructed by applying the same algorithm as used in the reconstruction of the simulated EASs (again demanding at least 2 MIPs in each detector). KASCADE reconstructed the direction of these EASs with zenith angles between $0^{\mathrm{o}}$ and $40^{\mathrm{o}}$ with $0.3^{\mathrm{o}}$ accuracy \cite{kascade2003}. Figure \ref{fig:directionreconstruction} shows the $1\,\sigma$ uncertainty on the HiSPARC direction reconstruction (red dots). The average uncertainty ($\theta < 40^{\mathrm{o}}$) is $6.1^{\mathrm{o}}$. Figure \ref{fig:reconstructionsimulations} gives the decomposition into the azimuth (red dots) and zenith (red crosses) angles as a function of the zenith angle.

When compared to KASCADE data it appears that the uncertainty obtained from the simulations ($7.7^{\mathrm{o}}$) slightly underestimates the real direction reconstruction performance of the HiSPARC station ($6.1^{\mathrm{o}}$). However, the simulated data only contain proton initiated showers with energies starting at $10^{13}$ eV. The KASCADE data contain all primary cosmic ray compositions while only showers with an energy in excess of $10^{14}$ eV are reconstructed. This results in a higher contribution from events with slightly higher particle densities. Moreover, in the simulations the shower core positions were evenly distributed in all directions up to a distance of 100 m from the station centre. The nearest boundaries of the (square) KASCADE array are in two directions only $\sim$55 m and $\sim$70 m away from the HiSPARC station \cite{proefschriftDavid}. The contribution from showers with their core position close to or beyond these boundaries are therefore suppressed. This will also reduce the number of lower multiplicity events observed by the HiSPARC station.

Recently, one month of data (April 2019) from four closely spaced four-detector diamond shaped stations (\#501, \#510, \#512 and \#513) was analysed \cite{internalJos}. The relative distance between the centers of the stations ranges from $\sim$1.5 m to $\sim$5 m. The direction of showers was reconstructed provided all four stations triggered at the same time and all 16 detectors observed a signal corresponding to at least 2 MIPs. These conditions favour the high multiplicity region in air showers and resulted in 123800 events. In each station the direction of the shower was reconstructed. By pairwise comparing the directions 6 distributions were obtained. For each distribution the $1\,\sigma$ difference between the two measurements was calculated. The results are listed in Table \ref{tab:adist}. They agree well with the outcome of the KASCADE data analysis.

\begin{table}
\centering
\begin{tabular}{l l}
\hline
\textbf{Stations} & \textbf{Difference [$\si{\degree}$]} \\
\hline
501-510  & $6.02\pm 0.04$ \\
501-512  & $6.14\pm 0.04$ \\
501-513  & $6.06\pm 0.04$ \\
510-512  & $6.35\pm 0.04$ \\
510-513  & $6.37\pm 0.04$ \\
512-513  & $5.93\pm 0.04$ \\
\hline
\end{tabular}
\caption{The first column lists the pairs of stations for which the reconstruction of the shower direction was compared. The second column gives the $1\,\sigma$ difference between the two angles. For details see \cite{internalJos}.}
\label{tab:adist}
\end{table}

\section{EAS energy reconstruction}
\label{sec:energyreconstruction}
Despite the evident limitations of a single HiSPARC station, i.e. the limited number of detectors, the relatively small distance between the detectors and the limited precision in the determination of the number of particles traversing a detector, it is possible to reconstruct the energy of the primary cosmic ray using a single four-detector station. The energy is determined by fitting a lateral density function (LDF), i.e. an analytical description of the particle densities in EAS footprints, to the measured particle densities. The number of particles traversing a detector is determined by dividing the measured pulse integral by the MIP-peak value. Since only the particle numbers at four detector positions can be determined, the maximum number of free parameters in the LDF is three; two for the core position and one for the energy. The core positions of low energy showers ($\sim$$10^{14.5}$ eV to $\sim$$10^{15.5}$ eV) are expected to lie within the station since EASs with their core outside the station are not likely to result in a trigger. For large showers there are two solutions to the LDF fitting procedure; one with the core inside and one outside the station area. Also, if all four detectors measure a similar number of particles, the data contain no information that can be used to accurately reconstruct the core position and thus the energy. Finally, in order to determine the energy flux, the effective surface area of an EAS footprint as function of energy needs to be known. Proton induced CORSIKA simulations in the energy range from $10^{14.5}$ eV to $10^{16.5}$ eV have been used to determine the effective surface area and to define the LDF.

\begin{figure}
	\centering
	\includegraphics[width=90mm]{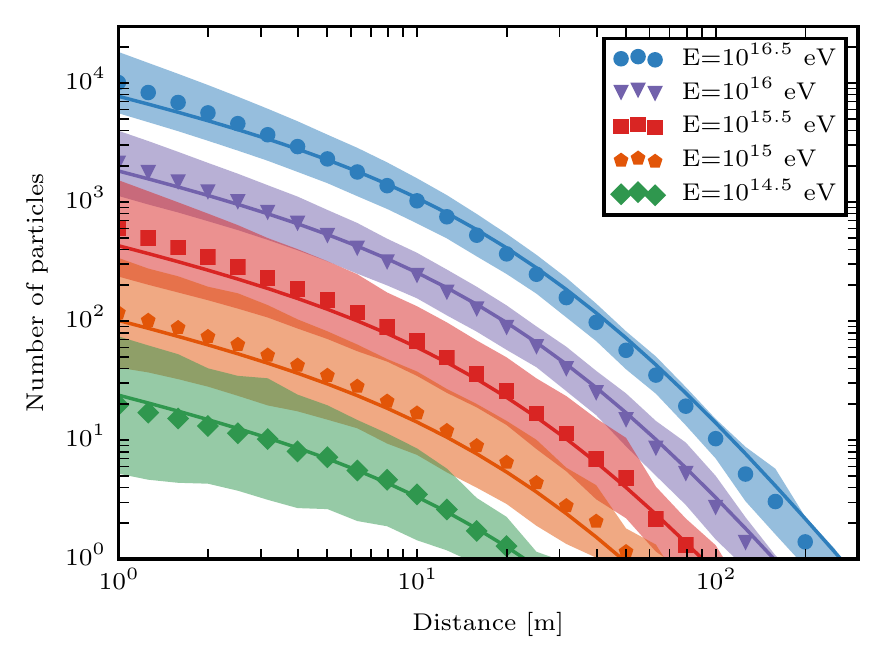}
	\caption{The simulated average number of particles that fall inside a 0.5 by 1 m rectangle as a function of distance to the shower core at different energies. The markers show the average particle numbers. The shaded regions show the 1$\sigma$ fluctuations. The lines show the modified NKG formula for each energy.}
	\label{fig:nkgfit}
\end{figure}

Figure \ref{fig:nkgfit} shows the simulated particle numbers as a function of distance to the shower core for multiple energies. For each shower the number of particles that traverse a 1 m by 0.5 m detector has been determined for various core positions. The markers display the averages of these particle numbers. For each energy a different marker is used. Only perpendicular incident proton simulations are shown. The fluctuation in the number of particles within showers of the same energy is large. The shaded regions show the 1$\sigma$ spread. The lines show the LDF used for the energy reconstruction. Despite the limited number of free parameters, the particle numbers are still well described by the function. A modification of the Nishimura-Kamata-Greisen formula \cite[and references therein]{gaisser2016} has been used as LDF. The number of free parameters in the original NKG formula has been reduced. The age parameter which is related to the shower maximum is effectively taken as a constant. In the simplified NKG formula the number of particles $N$ at distance $r$ is given by:
\begin{equation}
N(r) = A \left( \frac{r}{r_o} \right)^a \left( 1 + \frac{r}{r_o} \right)^b
\end{equation}
with $r_0=29.6$, $a=-0.566$, $b=-2.57$ and $A$ the fit parameter related to the energy. For inclined showers the particle numbers are reduced due to the increase in path length through the atmosphere. This is corrected for by using:
\begin{equation}
A_{\bot} = A \cdot \exp\left(p \left(\frac{1}{\cos \theta} - 1\right)\right)
\end{equation}
with $\theta$ the zenith angle and $p=6.937$. The energy of the primary cosmic ray is then calculated with:
\begin{equation}
\log(E) = c \cdot (\log(A_{\bot}) + d)
\end{equation}
here $c = 0.797$ and $d = 17.62$. The modified NKG formula is circle symmetric and two parameters are required to determine the core position.

\begin{figure}
	\centering
	\includegraphics[width=90mm]{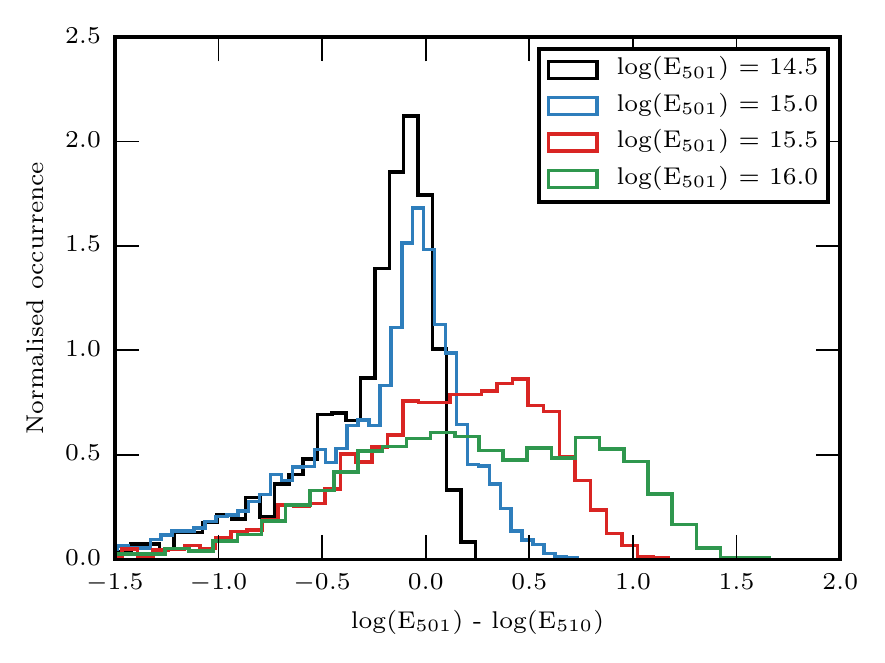}
	\caption{Distributions of reconstructed EAS energy differences between station 501 and 510 (colour online). Four distributions are displayed each containing a selection of energies as determined by station 501 (within a $\log(E)=0.5$ bin width). At higher energies a bimodal distribution appears reflecting the two solutions for the core position (inside and outside the station).}
	\label{fig:energyuncertainty}
\end{figure}

Three months of coincidences between stations 501 and 510 have been used to investigate the energy resolution obtained using the modified NKG method. Only events in which all 8 detectors of both stations measured a signal above 2 MIPs are considered. Furthermore, only events for which both directions deviate less than 15$\si{\degree}$ are used. Also, the average zenith angle has to be less than 35$\si{\degree}$. Finally, if the spread in the number of MIPs in the four detectors is less than 2, the event is discarded. This selection avoids events with similar particle numbers that contain no information about the core position. Coincident events between station 501 and 510 provide two independent energy measurements. Per reconstruction two initial guesses for the parameters are used. One with the core position inside the four detectors and another one outside. The fit parameters of the solution with the lowest $\chi^2$ value are used. If the best $\chi^2$ value of one of the two stations is below 5, the event is discarded. Figure \ref{fig:energyuncertainty} shows the distributions of reconstructed EAS energy differences between station 501 and 510. Four distributions are displayed each containing a selection of energies as determined by station 501 (within a $\log(E)=0.5$ bin width). At higher energies a bimodal distribution appears reflecting the two solutions for the core position. Multiple stations at sufficiently large distance are required to resolve this problem.

\begin{figure}
	\centering
	\includegraphics[width=90mm]{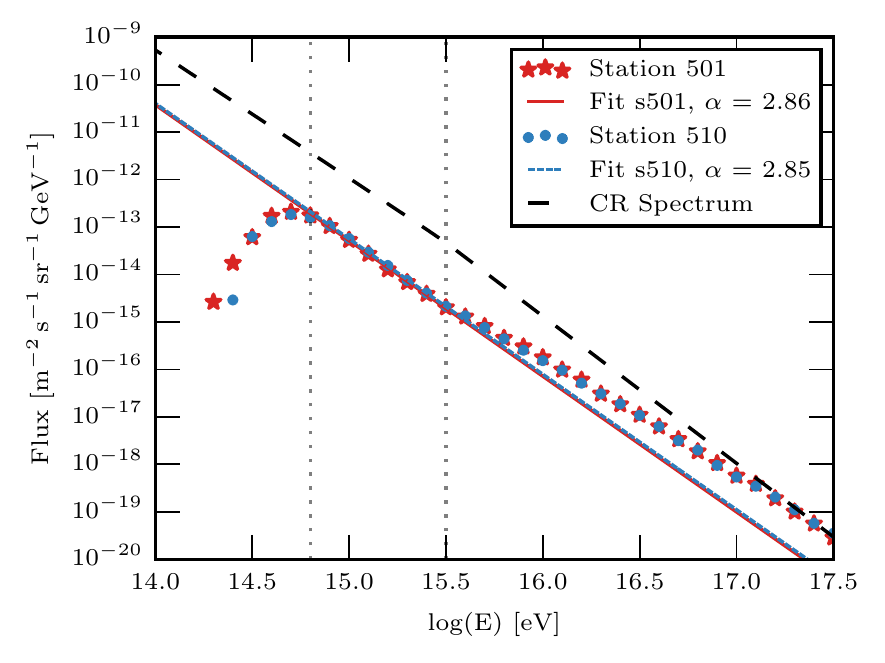}
	\caption{The energy spectrum obtained using stations 501 (red stars) and 510 (blue dots). For energies between $10^{14.8}$ eV and $10^{15.5}$ eV (dotted lines) a slope ($\alpha$) has been fitted to both distributions (solid lines). For higher energies the measured energy spectrum starts to soften whereas steepening is expected. This is due to the bimodal energy reconstruction for higher energies. The black dashed line shows the known energy spectrum. The large offset between the HiSPARC measurement and the spectrum is due to the detection efficiency and because of the numerous analysis cuts that are applied.}
	\label{fig:hisparcenergyspectrum}
\end{figure}

The cosmic ray energy spectrum is obtained by dividing the distribution of energy measurements by the solid angle (zenith angle of 35$\si{\degree}$), the time span (3 months) and the effective surface area of an EAS footprint. CORSIKA simulations have been used to obtain the radius at which the particle number drops below 2 (Fig. \ref{fig:nkgfit}). The effective surface area at an energy is determined as the area of a circle with that radius. The radii at intermediate energies have been estimated by fitting a second order polynomial to the radii of the simulated energies. Figure \ref{fig:hisparcenergyspectrum} shows the energy spectrum obtained using stations 501 (red stars) and 510 (blue dots). Both stations yield very similar results. For energies between $10^{14.8}$ eV and $10^{15.5}$ eV (dotted lines) a slope ($\alpha$) has been fitted to both distributions (solid lines). The values of these slopes (2.85 and 2.86) are similar and do not deviate much from the known value of 2.7 \cite{gaisser2016}. For higher energies the measured energy spectrum starts to soften whereas a steepening is expected. This follows from the bimodal distribution at higher energies (Fig. \ref{fig:energyuncertainty}). If only solutions to the energy reconstruction problem with the core position inside the four detectors are taken into account, the spectrum steepens rapidly and no energies larger than $10^{16}$ are measured. The black dashed line shows the known energy spectrum with a steepening to a slope of 3.1 at $3\cdot10^{15}$ eV. The large offset between the HiSPARC measurement and the spectrum is due to the detection efficiency and because of the numerous analysis cuts that are applied (e.g. all detectors have at least 2 MIPs).

A similar analysis using KASCADE data has been carried out. Unfortunately the number of KASCADE reconstructions was too limited for a decisive analysis. The uncertainty in the $\sim$$10^{15}$ eV region seems to agree with the analysis using stations 501 and 510. For two-detector stations it is not possible to reconstruct the energy of individual EASs. However, by investigating the pulse height distribution it is possible to probe the cosmic ray energy spectrum. This will be discussed in a separate paper \cite{kasper2019}. In several regions stations are positioned close enough that a combination of stations can be used for energy reconstruction. A preliminary study with the Science Park Cluster \cite[Chapter~7]{proefschriftHans} has been carried out. A more elaborate study that uses an AI algorithm in combination with properties derived from the pulse size and pulse shape (e.g. particle multiplicity and arrival time) is in progress \cite{kasper2019}.

\section{Shower detection}
\label{sec:showerdetection}
To derive the EAS detection efficiency of a station a detailed comparison has to be made between the measured detector response curve and the pulse height distribution obtained in the simulation. Showers are again generated in an energy range from $10^{13}$ to $10^{16.5}$ eV following the cosmic ray energy spectrum with randomly chosen directions and EAS core positions.

\subsection{Pulse height distribution}
\begin{figure}
\centering
\includegraphics[width=90mm]{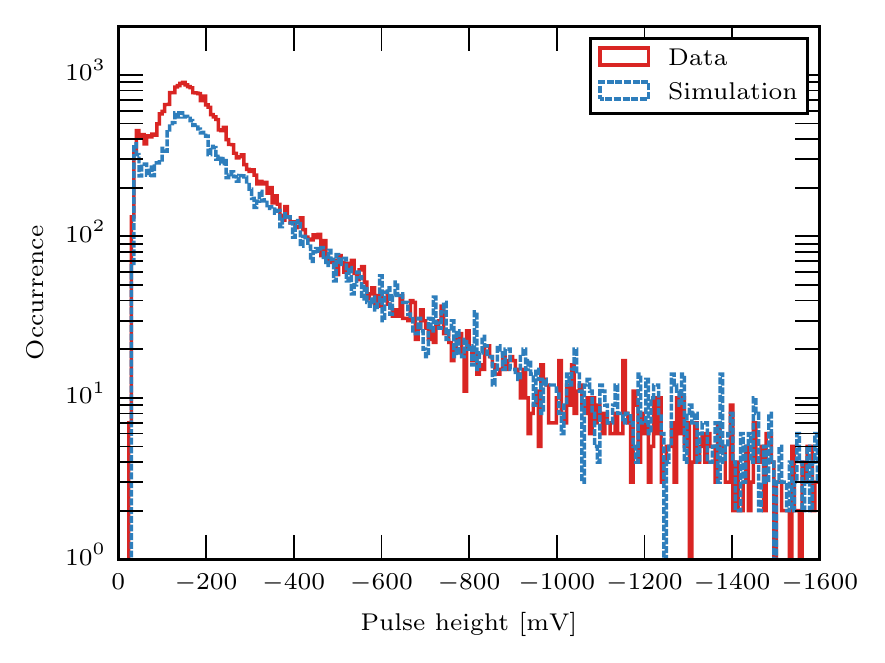}
\caption{Pulse height distribution of a detector in a two-detector station (red, solid) compared to simulations (blue, dashed). The contribution from random coincidences has been removed from the data while the MIP-peak value has been corrected for temperature fluctuations. For large pulses, data and simulations agree rather well. The discrepancy for small pulses is caused by the absence of muon induced 'mini-showers' (muons decaying into single electrons which in turn generate an  electromagnetic shower) in the simulations. These events are caused by the large number of showers for which the primary particle carries an energy less than $\sim$$10^{12.5}$ eV.}
\label{fig:pulseheights-comparison}
\end{figure}

Figure \ref{fig:pulseheights-comparison} shows the pulse height distribution of a detector in a two-detector station compared to simulations. For pulses in excess of $-400$ mV the simulation (blue) agrees reasonably well with the data (red). Note that EASs with higher energies than $10^{16.5}$ eV are not included. For small pulses however, there appears to be a clear discrepancy. In the simulation however, an important contribution is absent. Only EASs with energies larger than $10^{13}$ eV are considered. Lower energy showers responsible for the single muon background are not taken into account. When an energetic muon decays prior to reaching the Earth it will produce an energetic electron. This electron initiates an electromagnetic 'mini-shower'. These muon-induced mini-showers are indeed present in for instance $10^{10}$ eV proton induced CORSIKA showers. Adding this contribution resolves the discrepancy and is discussed in detail in \cite{kasper2019}.

\begin{figure}
\centering
\includegraphics[width=90mm]{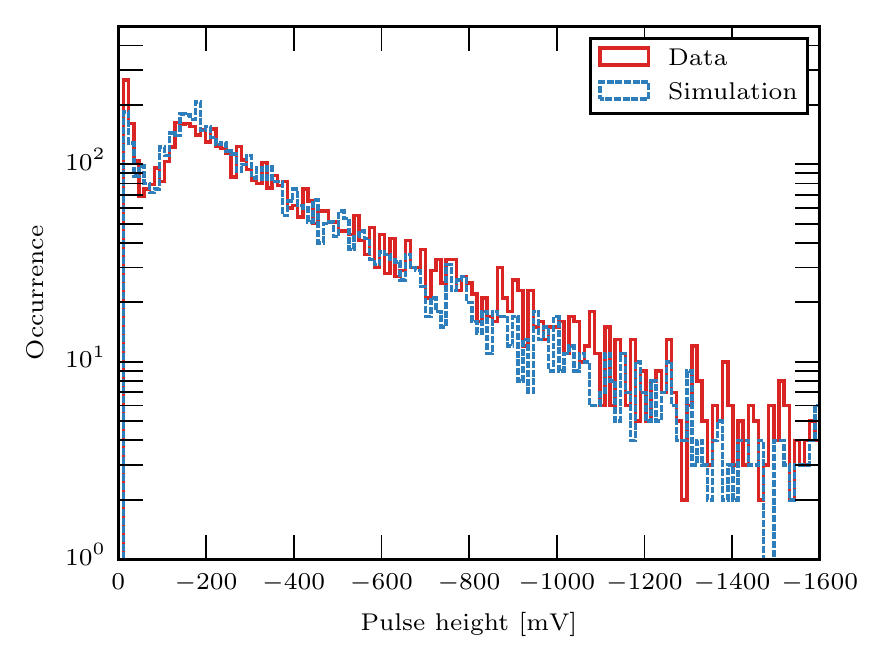}
\caption{Pulse height distribution of a detector in a four-detector station (red, solid) compared to simulations (blue, dashed). Only events that contain a signal in excess of $-30$ mV in each of the four detectors are selected. In contrast to Figure \ref{fig:pulseheights-comparison}, there is no discrepancy between data and simulation for small pulse heights as the size of the shower footprint has to be as large as or larger than the size of the station (i.e. EASs with an energy larger than $\sim$$10^{13}$ eV). The discrepancy at large pulse heights becomes apparent; the simulations do not include EASs with energies larger than $10^{16.5}$ eV.}
\label{fig:pulseheights-4d}
\end{figure}

Alternatively, contributions from muon-induced mini-showers can also be rejected. As the number density of these mini-showers is low, the probability of detecting a mini-shower rapidly decreases when three or more detectors are required to produce a signal in excess of $-30$ mV. Figure \ref{fig:pulseheights-4d} shows the simulated pulse height distribution for a detector in a four-detector station (blue) and the measured pulse height spectrum (red). All four detectors generated a signal in excess of $-30$ mV. The discrepancy for small pulse heights completely disappears; demanding a signal in all four detectors completely removes contributions from single muon induced mini-showers. Contrary to what is observed in Figure \ref{fig:pulseheights-comparison}, the experimental data now show a pronounced excess for large pulses. Since EAS energies beyond $10^{16.5}$ eV are not included, indeed the simulations underestimate the contribution from EASs with higher particle densities \cite{kasper2019}.

\begin{figure}
\centering
\includegraphics[width=90mm]{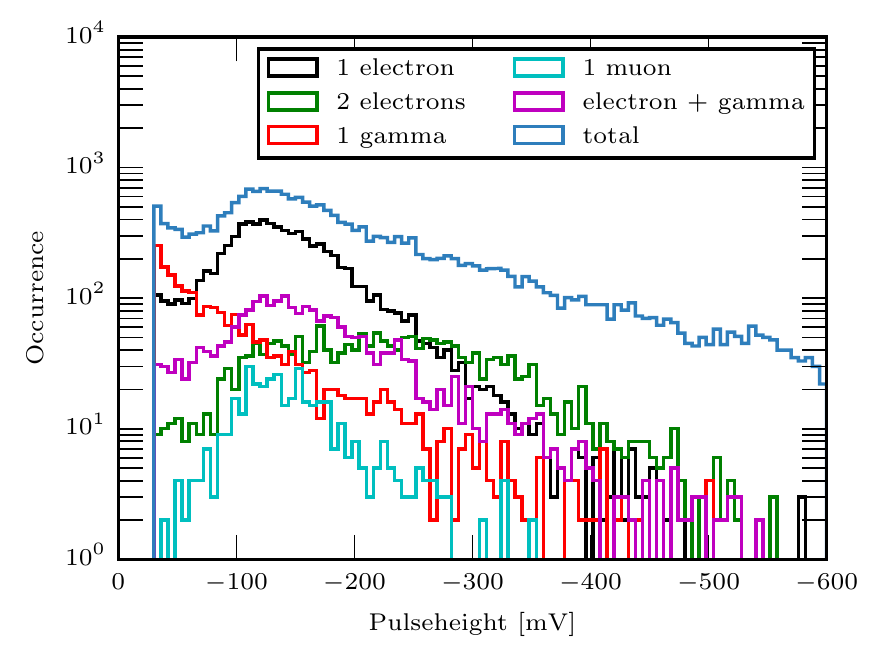}
\caption{Simulated response (blue histogram) of a detector in a two-detector station (colour online). The distribution receives contributions from electrons (single electrons: black, 2 electrons: green), single muons (light blue), single gamma rays (red), single electron plus single gamma (purple) and higher multiplicities of various combinations (not shown separately). This simulation can be used to attribute a likelihood to finding the particle multiplicity in a detector at a specific pulse height.}
\label{fig:pulseheights}
\end{figure}

The pulse height distributions in Figures \ref{fig:pulseheights-comparison} and \ref{fig:pulseheights-4d} receive contributions from electrons, muons and gamma rays. This is demonstrated in Figure \ref{fig:pulseheights} where the contribution from a number of particles (and their combinations) to the simulated spectrum in Figure \ref{fig:pulseheights-comparison} (blue) is shown. The contributions from one electron (black), one gamma (red), one muon (light blue), two electrons (green) and electron plus gamma (purple) are shown separately. Small pulses are predominantly generated by single gamma rays and single low energy electrons. The blue histogram is the sum of all these contributions and also includes higher multiplicity combinations; it matches the simulated pulse height distribution in Figure \ref{fig:pulseheights-comparison}. Figure \ref{fig:pulseheights} also demonstrates that the pulse height is a measure for the number of particles that traverses a detector.

\subsection{EAS detection efficiency}
The probability for detecting an EAS with a single two-detector station as function of distance between the shower core and station center and zenith angle is investigated with CORSIKA shower simulations. The station was exposed to EASs with an energy of $10^{15}$ eV. The direction of primary cosmic rays is chosen uniformly between $0^{\mathrm{o}}$ and $60^{\mathrm{o}}$. The positions of the shower core with respect to the station center were homogeneously distributed within a circle with a radius of 150 m. Note that for increasing zenith angle the size of the shower footprint augments, while at the same time the chance of particle absorption in the atmosphere significantly increases. Both effects result in a lower particle density in the footprint. Also, the number of EASs that arrive at large core distances is larger than at small core distances because of the homogeneous exposure and the increasing effective area at larger (ring shaped) core distance bins. The same goes for the zenith angle. The solid angle $\Omega$ of the circular field of view subtended by a rotated zenith angle $\theta$ is given by $\Omega = 2\pi(1-\cos(\theta))$. This implies that larger zenith angles result in larger solid angles.

\begin{figure}
\centering
\includegraphics[width=90mm]{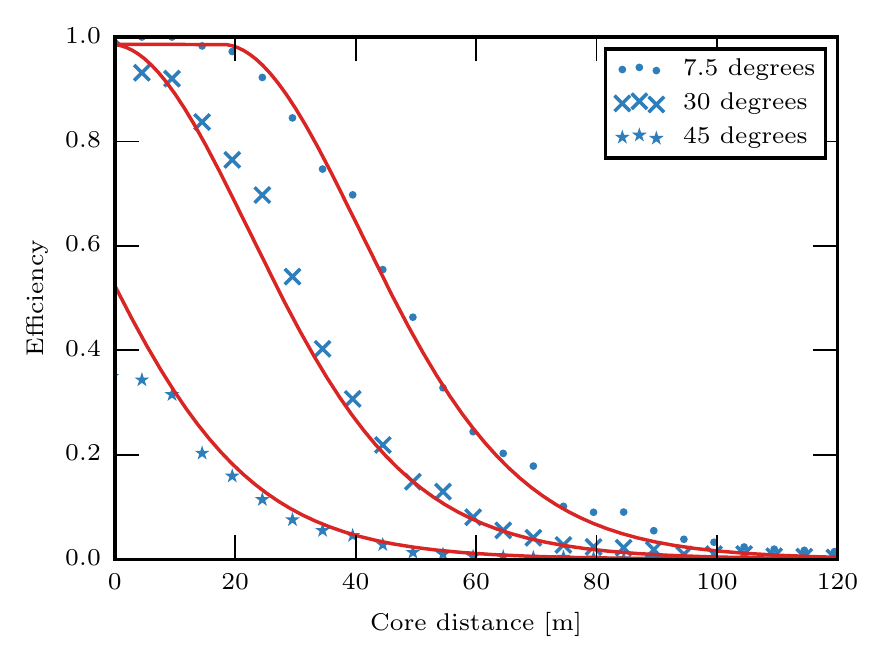}
\caption{EAS detection efficiency of $10^{15}$ eV showers as function of distance between the shower core and station center for zenith angles of $7.5^{\mathrm{o}}$ (dots), $30^{\mathrm{o}}$ (crosses) and $45^{\mathrm{o}}$ (stars). At $7.5^{\mathrm{o}}$ and for small core distances the efficiency is close to 1. For larger core distances the EAS detection efficiency decreases rapidly. The lines display the parametrisation in eq. \ref{eq:emg}.}
\label{fig:efficiency_coredistance}
\end{figure}

\begin{figure}
\centering
\includegraphics[width=90mm]{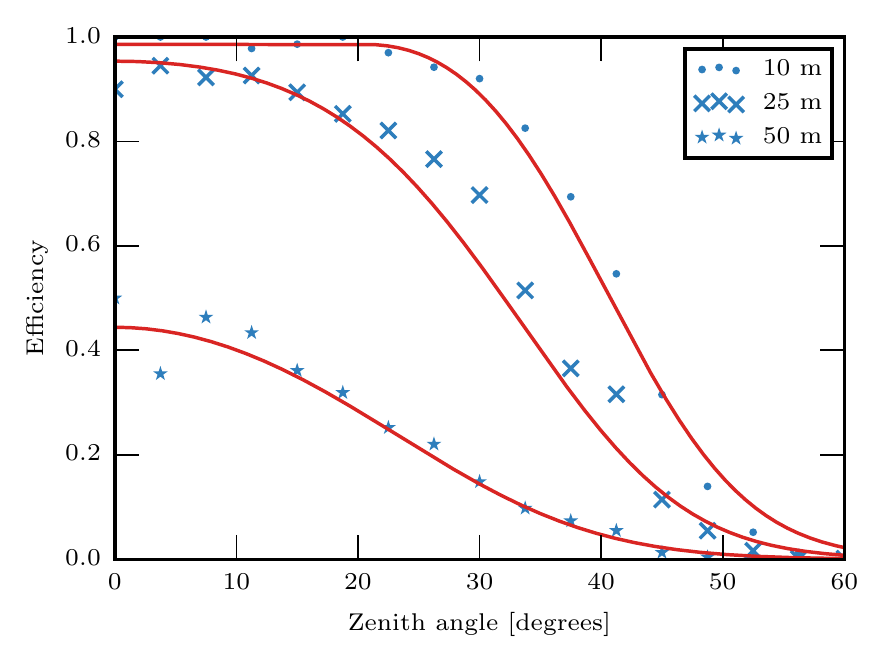}
\caption{EAS detection efficiency of $10^{15}$ eV showers as function of zenith angle for core distances of 10 (dots), 25 (crosses) and 50 m (stars). If the shower core is close to the center of the station (e.g. 10 m in the figure) the EAS detection efficiency stays close to 1 up to relatively large zenith angles. The lines display the parametrisation in eq. \ref{eq:emg}.}
\label{fig:efficiency_zenith}
\end{figure}

Figure \ref{fig:efficiency_coredistance} shows the EAS detection efficiency as function of core distance for zenith angles of $7.5^{\mathrm{o}}$ (blue dots), $30^{\mathrm{o}}$ (blue crosses) and $45^{\mathrm{o}}$ (blue stars). At $7.5^{\mathrm{o}}$ and for small core distances the efficiency is close to 100\%. With increasing core distance the EAS detection efficiency rapidly decreases. Figure \ref{fig:efficiency_zenith} shows the detection efficiency as function of zenith angle for core distances of 10 (blue dots), 25 (blue crosses) and 50 m (blue stars). If the shower core is close to the center of the station (e.g. 10 m in the figure) the EAS detection efficiency remains close to 100\% for even relatively large zenith angles. If the shower core is further away from the station center the efficiency becomes much lower.

A 2D parametrisation (combining the fits in Figures \ref{fig:efficiency_coredistance} and \ref{fig:efficiency_zenith}) is derived that describes the detection efficiency as a function of the distance between station and shower core, and zenith angle. For small core distances, because of the high number density near the shower core, the EAS detection efficiency as function of core distance is expected to be 100\%. At a distance $r_m$, at which the probability for EAS particles to miss a detector becomes substantial (e.g. for the $7.5^{\mathrm{o}}$ EAS detection efficiency in Fig. \ref{fig:efficiency_coredistance} the value of $r_m$ is $\sim$20 m), the efficiency decreases. For core distances larger than $r_m$ the radial dependency of the EAS detection efficiency can accurately be described by the formula of an exponentially modified Gaussian distribution. The zenith angle dependence is obtained by shifting $r_m$ to smaller and eventually negative values depending on the zenith angle. This leads to the following parametrisation:
\begin{equation}
\label{eq:emg}
p(\boldsymbol{r},\boldsymbol{\theta})=
\begin{cases}
f(r_m, \alpha, \mu(\boldsymbol{\theta}, \chi, \rho), \sigma, \lambda)&  \textup{for $r < r_m$}\\
f(\boldsymbol{r}, \alpha, \mu(\boldsymbol{\theta}, \chi, \rho), \sigma, \lambda)&  \text{for $r \geq r_m$}
\end{cases}
\end{equation}
with the modified Gaussian distribution $f(r, \alpha, \mu, \sigma, \lambda)$ given by:
\begin{align}
f(r, \alpha, \mu, \sigma, \lambda) = & \quad \alpha \exp \left[\frac{\lambda}{2}(2\mu + \lambda \sigma^2 - 2r)\right] \nonumber\\
&\times \mathrm{erfc}\left( \frac{\mu + \lambda \sigma^2 -r}{\sqrt{2}\sigma} \right)
\end{align}
$\mu$ and $\sigma$ are the mean and standard deviation of the Gaussian part of the distribution and $\lambda$ is the rate of the exponential distribution. The $\mathrm{erfc}(x)$ factor is the complementary error function and is given by:
\begin{equation}
\textup{erfc}(x) = \frac{2}{\sqrt{\pi}}\int_x^\infty e^{-y^2}dy
\end{equation}
The value of $r_m$ is the mode of $f(r, \alpha, \mu, \sigma, \lambda)$ and depends on $\mu$, $\sigma$ and $\lambda$. The function $f(r)$ is thus continuous at $r_m$. The shift of $r_m$ as function of the zenith angle depends on $\mu$. The zenith angle dependency is thus absorbed in $\mu$:
\begin{equation}
\mu(\theta, \chi, \rho) = (\chi + \rho) \exp [-(\sec \theta -1)] - \chi
\end{equation}
The exponential decrease of $\mu$ (and thus $r_m$) with the secant of the zenith angle can be thought of as an exponential diminishing of the shower particle density for larger travel distances through the atmosphere.

\begin{figure}
\centering
\includegraphics[width=90mm]{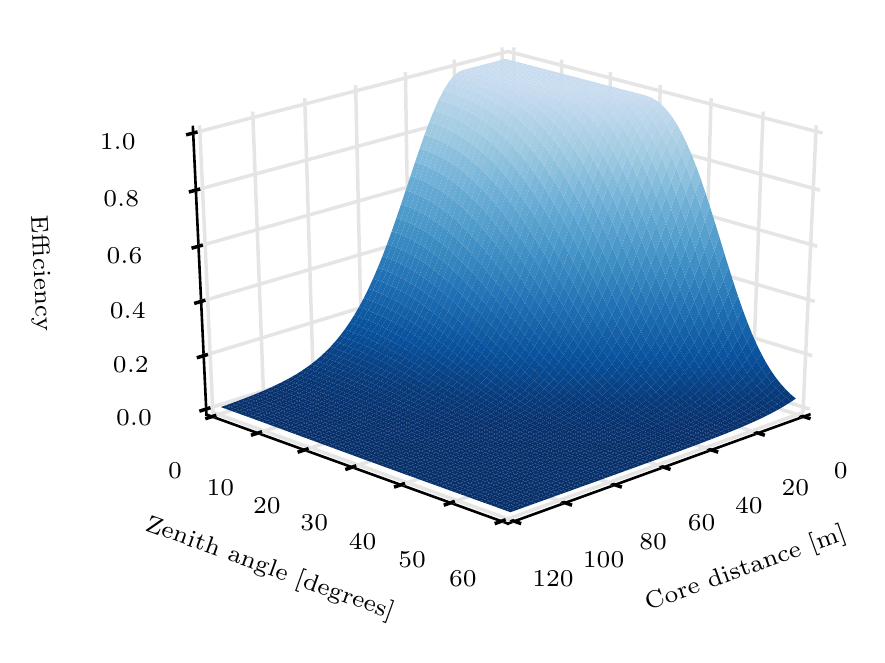}
\caption{A 3D display of the parametrised EAS detection efficiency of $10^{15}$ eV proton induced showers. For small zenith angles and core distances the efficiency is close to 1. For larger angles and distances the efficiency decreases but the surface area and solid angle increase substantially.}
\label{fig:2dfit}
\end{figure}

The lines in Figures \ref{fig:efficiency_coredistance} and \ref{fig:efficiency_zenith} show the parametrisations obtained by fitting $\alpha$, $\sigma$, $\lambda$, $\rho$ and $\chi$ to the simulated EAS detection efficiency. The obtained fit parameters are 2.15, 20.9, $7.22 \cdot 10^{-2}$, 7.84 and 129 resp. Figure \ref{fig:2dfit} shows a 3D plot of the parametrised EAS detection efficiency. With increasing EAS energy the 'plateau' at $\sim$1 for small zenith angles and core distances increases. The shower detection efficiencies at other primary energies are discussed in \cite{kasper2019}.

\section{Summary and Conclusion}
\label{sec:conclusion}
The High School Project on Astrophysics Research with Cosmics (HiSPARC) is a large air shower detector array with approximately 140 stations throughout the Netherlands, United Kingdom, Denmark and Namibia. HiSPARC is a collaboration of universities, scientific institutes and high schools each hosting one or more detection stations. The majority is managed by high schools.

The detection principle of HiSPARC is to sample EAS footprints using scintillation detectors. A HiSPARC detector consists of a scintillator glued to a light-guide which is connected to a PMT. The single muon detector response shows good agreement with simulations.

A detector is placed inside a roof box on the rooftop of a building. A detection station consists of two or four detectors. Custom made electronics read out and control two PMTs. For four-detector stations two units are used in a Master-Slave configuration. Multiple stations can be combined to sample an EAS footprint. To precisely correlate coincident samples accurate absolute timing is required. All stations are equipped with a GPS receiver that provides location and the timestamp for an event.

In 2008 a HiSPARC four-detector station was integrated in the KASCADE experiment. The shower direction reconstruction from a HiSPARC station (uncertainty $\sim$$6^{\mathrm{o}}$) was compared to KASCADE measurements. Similar results were obtained by comparing the direction reconstruction of four closely spaced four-detector station sampling the same shower. CORSIKA EAS simulations have been used to verify the uncertainty as well. Measurements and simulation yield comparable results. CORSIKA was also applied to determine the EAS detection efficiency of a station as function of distance to the shower core and zenith angle of the primary cosmic ray.

For small showers of which the core position falls within the area covered by the HiSPARC station, the energy can be estimated. Combining more stations at distances up to $\sim$1 km, core reconstructions for larger showers improves considerably.

A subset of stations not too far from each other (distances smaller than the footprint size of the highest energy showers) forms a HiSPARC cluster. The Amsterdam Science Park Cluster is the densest cluster containing thirteen four-detector stations. Timing offsets between GPS antennas have been measured using EAS data. The field-of-view of the Science Park stations reaches beyond Polaris.

The large distance between stations in the network (up to 1000 km, excluding the station in Namibia) allows for searching coincident showers resulting from spallation of a primary cosmic ray (GZ-mechanism \cite{gz1960}). The HiSPARC array is also used to investigate the relation between cosmic rays and weather phenomena such as lightning \cite{JosBliksemNTvN}.

HiSPARC is collecting data for already more than 15 years. A Dutch high school curriculum containing chapters on quantum mechanics, particle physics and special relativity in combination with a low cost (5,000 \euro/10,000 \euro), robust HiSPARC station, easy accessible (pre-processed and raw) data and an extensive library with analyses tools, are the keys to success. HiSPARC brings science and research into the class room!

\section*{Acknowledgements}
The HiSPARC experiment could not have been realised without the interest of and contributions from the high schools, their physics teachers, technical assistants, ICT departments and, most importantly, their students. Schools managed to access public funds, but also received financial contributions from individuals and commercial businesses. The exhaustive list of persons, institutions and companies that (financially and/ or in person) support the project is too long to give here.

In 2004 HiSPARC received the 'Altran Foundation for Innovation' award 'Discovering, understanding and enjoying science through innovation' \cite{nature2004}, \cite{altran}. We are indebted to Altran for their commitment and the exposure the award generated.

The Dutch National Institute for Subatomic Physics (Nikhef) has been indispensable for the development and realisation of the project. The institute offered financial support, lab space, enabled the development of detector hardware and electronics, software and ICT infrastructure for data transfer and data storage. Without the generous help of Nikhef's former director Frank Linde and Nikhef's managing director Arjen van Rijn the realisation of the project would not have been possible.

We are grateful that the KASCADE collaboration offered us the opportunity to install a test station at their site in Karlsruhe, and supplied us with the necessary infrastructure and KASCADE data.



\bibliographystyle{elsarticle-num}
\bibliography{references.bib}

\begin{thebibliography}{10}
\expandafter\ifx\csname url\endcsname\relax
  \def\url#1{\texttt{#1}}\fi
\expandafter\ifx\csname urlprefix\endcsname\relax\def\urlprefix{URL }\fi
\expandafter\ifx\csname href\endcsname\relax
  \def\href#1#2{#2} \def\path#1{#1}\fi

\bibitem{gaisser2016}
T.~K. Gaisser, R.~Engel, E.~Resconi, {Cosmic rays and particle physics},
  Cambridge University Press, 2016.

\bibitem{hundhausen2012}
A.~J. Hundhausen, {Coronal expansion and solar wind}, Vol.~5, Springer Science
  \& Business Media, 2012.

\bibitem{sepreview2017}
K.-L. Klein, S.~Dalla, {Acceleration and propagation of solar energetic
  particles}, Space Science Reviews 212 (2017) 1107--1136.

\bibitem{blasi2013}
P.~Blasi, {The origin of galactic cosmic rays}, The Astronomy and Astrophysics
  Review 21 (2013) 70.

\bibitem{auger2017}
A.~Aab, P.~Abreu, M.~Aglietta, et~al., {Observation of a large-scale anisotropy
  in the arrival directions of cosmic rays above $8 \times 10^{18}$ eV},
  Science 357 (2017) 1266--1270.

\bibitem{grieder2010}
P.~K.~F. Grieder, {Extensive Air Showers: High Energy Phenomena and
  Astrophysical Aspects - A Tutorial, Reference Manual and Data Book}, Springer
  Science \& Business Media, 2010.

\bibitem{kascade2003}
T.~Antoni, W.~D. Apel, F.~Badea, et~al., {The cosmic-ray experiment KASCADE},
  Nucl. Instrum. Methods Phys. Res. A 513 (2003) 490--510.

\bibitem{apel2010}
W.~D. Apel, J.~C. Arteaga, A.~F. Badea, et~al., {The KASCADE-Grande
  experiment}, Nucl. Instrum. Methods Phys. Res. A 620 (2010) 202--216.

\bibitem{agasa1992}
N.~Chiba, K.~Hashimoto, N.~Hayashida, et~al., {Akeno giant air shower array
  (AGASA) covering 100 km2 area}, Nucl. Instrum. Methods Phys. Res. A 311
  (1992) 338--349.

\bibitem{sdtelescopearray2012}
T.~Abu-Zayyad, R.~Aida, M.~Allen, et~al., {The surface detector array of the
  Telescope Array experiment}, Nucl. Instrum. Methods Phys. Res. A 689 (2012)
  87--97.

\bibitem{auger2015}
{Pierre Auger Collaboration}, {The Pierre Auger cosmic ray observatory}, Nucl.
  Instrum. Methods Phys. Res. A 798 (2015) 172--213.

\bibitem{hisparcwebsite}
{HiSPARC website}, \url{https://www.hisparc.nl}.

\bibitem{proefschriftDavid}
D.~B. R.~A. Fokkema, {The HiSPARC Experiment}, Ph.D. thesis, University of
  Twente, {ISBN: 978-90-365-3438-3},
  \url{https://www.nikhef.nl/pub/services/biblio/theses_pdf/thesis_D_Fokkema.pdf}
  (2012).

\bibitem{nikhefwebsite}
{Nikhef website}, \url{https://www.nikhef.nl}.

\bibitem{laas1999}
T.~Wada, N.~Ochi, T.~Kitamura, et~al., {Observation of time correlation in
  cosmic air shower network}, in: {Nuclear Physics B - Proceedings
  Supplements}, Vol.~75, 1999, pp. 330--332.

\bibitem{hofverberg2006}
P.~Hofverberg, {Imaging the high energy cosmic ray sky}, Ph.D. thesis, KTH
  (2006).

\bibitem{chicos2009}
E.~Brobeck, {Measurement of ultra-high energy cosmic rays with CHICOS}, Ph.D.
  thesis, California Institute of Technology (2009).

\bibitem{eee2013}
M.~Abbrescia, A.~Agocs, S.~Aiola, et~al., {The EEE experiment project: status
  and first physics results}, The European Physical Journal Plus 128 (2013) 62.

\bibitem{timmermans2003}
C.~Timmermans, J.~Schotanus, B.~van Eijk, et~al., {Wetenschappelijk onderzoek
  voor scholieren}, Nederlands Tijdschrift voor Natuurkunde 69 (2003) 230--234.

\bibitem{bc408datasheet}
{BC408 Data Sheet}, Saint-Gobain Ceramics \& Plastics, Inc (2016).

\bibitem{pmtelectrontubes}
{9107B series data sheet}, ET Enterprises Ltd (2010).

\bibitem{pmthamamatsu}
{R6094 data sheet}, Hamamatsu (1996).

\bibitem{landau1944}
L.~Landau, {On the energy loss of fast particles by ionization}, J. Phys.(USSR)
  8 (1944) 201--205.

\bibitem{baseelectrontubes}
{HV3020 series data sheet}, ET Enterprises Ltd (2018).

\bibitem{nikhefbase}
P.~Timmer, E.~Heine, H.~Peek, {Very low power, high voltage base for a Photo
  Multiplier Tube for the {KM}3NeT deep sea neutrino telescope}, Journal of
  Instrumentation 5 (2010) C12049--C12049.

\bibitem{GEANT42016}
J.~Allison, K.~Amako, J.~Apostolakis, et~al., {Recent developments in Geant4},
  Nucl. Instrum. Methods Phys. Res. A 835 (2016) 186--225.

\bibitem{kasper2019}
K.~van Dam, et~al., in preparation.

\bibitem{dayabay2012}
S.~Jetter, D.~Dwyer, J.~Wen-Qi, et~al., {PMT waveform modeling at the Daya Bay
  experiment}, Chinese Physics C 36 (2012) 733.

\bibitem{reyna2006}
D.~Reyna, {A Simple Parameterization of the Cosmic-Ray Muon Momentum Spectra at
  the Surface as a Function of Zenith Angle}, arXiv preprint:
  arXiv:hep-ph/0604145.

\bibitem{hastings1970}
W.~K. Hastings, {Monte Carlo sampling methods using Markov chains and their
  applications}, Biometrika 57 (1970) 97--109.

\bibitem{corsika1998}
D.~Heck, J.~Knapp, J.~N. Capdevielle, et~al., {CORSIKA: a Monte Carlo code to
  simulate extensive air showers.}, Forschungszentrum Karlsruhe GmbH, Karlsruhe
  (Germany), 1998.

\bibitem{DavisVantagePro}
{Davis Vantage Pro}, Davis Instruments, see
  \url{https://www.davisinstruments.com}.

\bibitem{TimKokkeler}
T.~Kokkeler, Improvement of the pulse reconstruction and arrival direction
  estimation at the hisparc experiment, {Bachelor thesis} (2019).

\bibitem{gpsmodule}
{Resolution T GPS Embedded Board User Guide}, Trimble (2009).

\bibitem{labview2007}
C.~Elliott, V.~Vijayakumar, W.~Zink, et~al., {National instruments LabVIEW: a
  programming environment for laboratory automation and measurement}, JALA:
  Journal of the Association for Laboratory Automation 12 (2007) 17--24.

\bibitem{mysql}
{MySQL Reference Manual}, Oracle Corporation, available at
  \url{https://www.mysql.com}.

\bibitem{python}
{Python Language Reference, version 2.7}, Python Software Foundation, available
  at \url{https://www.python.org}.

\bibitem{nagios}
{Nagios Core Documentation}, Nagios Enterprises, LLC, available at
  \url{https://www.nagios.org}.

\bibitem{openvpn}
{OpenVPN Documentation}, OpenVPN Inc., available at
  \url{https://www.openvpn.net}.

\bibitem{tightvnc}
{TightVNC Documentation}, GlavSoft LLC, available at
  \url{https://tightvnc.com}.

\bibitem{wilson1956}
J.~G. Wilson, {Progress in Cosmic Ray Physics}, Vol.~3, North-Holland
  Publishing Company, 1956.

\bibitem{NiekArxiv}
N.~G. Schultheiss, {The acceptance of the HiSPARC experiment}, arXiv preprint:
  arXiv:1602.06799.

\bibitem{hdf5}
{The HDF5 library \& file format}, The HDF Group, available at
  \url{https://www.hdfgroup.org/solutions/hdf5}.

\bibitem{django}
{Django Documentation}, Django Software Foundation, available at
  \url{https://www.djangoproject.com}.

\bibitem{sapphire2012}
D.~B. R.~A. Fokkema, A.~P. L.~S. de~Laat, T.~Kooij, {SAPPHiRE},
  \url{https://github.com/HiSPARC/sapphire} (2012).

\bibitem{jSparc}
{jSparc website}, \url{https://docs.hisparc.nl/jsparc}.

\bibitem{jupyternotebook}
{Jupyter Notebook Documentation}, Jupyter Team, available at
  \url{https://www.jupyter.org}.

\bibitem{hillas1981}
A.~M. Hillas, {}, in: Proceedings International Cosmic Ray Conference, Vol.~1,
  1981, p. 193.

\bibitem{qgsjet2006}
S.~Ostapchenko, {QGSJET-II: towards reliable description of very high energy
  hadronic interactions}, in: Nuclear Physics B - Proceedings Supplements, Vol.
  151, 2006, pp. 143--146.

\bibitem{gheisha1985}
H.~Fesefeldt, {GHEISHA The Simulation of Hadronic Showers}, Tech. rep., RWTH
  Aachen, {PITHA-85/02} (1985).

\bibitem{egs41985}
W.~R. Nelson, D.~W.~O. Rogers, H.~Hirayama, {The EGS4 code system}, Tech. rep.,
  Stanford Linear Accelerator Center, Stanford, California (1985).

\bibitem{proefschriftHans}
J.~M.~C. Montanus, {The Observability of Jets in Cosmic Air Showers}, Ph.D.
  thesis, University of Amsterdam, {ISBN: 978-94-028-0549-9},
  \url{https://www.nikhef.nl/pub/services/biblio/theses_pdf/thesis_H_M_Montanus.pdf}
  (2017).

\bibitem{internalJos}
J.~J.~M. Steijger, Het kwartet, {Internal note} (2019).

\bibitem{gz1960}
N.~M. Gerasimova, G.~T. Zatsepin, {Disintegration of cosmic ray nuclei by solar
  photons}, JETP 38 (1960) 1245--1252.

\bibitem{JosBliksemNTvN}
R.~Beekman, T.~Aldus, J.~Steijger, {Speelt kosmische straling een rol bij
  initiatie bliksem?}, Nederlands Tijdschrift voor Natuurkunde 84 (2018) 18.

\bibitem{nature2004}
{Schools at $10^{20}$ eV and beyond}, Nature 429 (2004) 685.

\bibitem{altran}
{2004 Altran award summary},
  \url{http://www.altran-foundation.org/fileadmin/medias/1.fondation/documents/Foundation_2004_Award.pdf}.

\end{thebibliography}

\end{document}